\documentclass{aa}  

\usepackage{graphicx}
\usepackage{txfonts}
\usepackage[]{natbib}
\usepackage{ amssymb }
\usepackage{ gensymb }
\bibpunct{(}{)}{;}{a}{}{,}
\begin{document}

   \title{Spectroscopy of the short-hard GRB\,130603B}
   \subtitle{The host galaxy and environment of a compact object merger}

   \author{
   A. de Ugarte Postigo\inst{1,2,*}
   \and
   C.C. Th\"one\inst{1}
   \and 
   A. Rowlinson\inst{3}
   \and
   R. Garc\'ia-Benito\inst{1}
   \and
   A.J. Levan\inst{4}
   \and
   J. Gorosabel\inst{1,5,6}
   \and
   P. Goldoni\inst{7}
   \and
   S. Schulze\inst{8,9}
   \and   
   T. Zafar\inst{10}
   \and 
   K. Wiersema\inst{11}
   \and 
   R. S\'anchez-Ram\'irez\inst{1}
   \and 
   A. Melandri\inst{12}
   \and 
   P. D'Avanzo\inst{12}
   \and 
   S. Oates\inst{13}
  \and 
   V.D'Elia\inst{14,15}
   \and 
   M. De Pasquale\inst{13}
   \and 
   T. Kr\"uhler\inst{2,16}
   \and 
   A. J. van der Horst\inst{3}
   \and 
   D. Xu\inst{2}
   \and 
   D. Watson\inst{2}
   \and 
   S. Piranomonte\inst{15}
   \and 
   S.D.~Vergani\inst{17,12}
   \and 
   B.~Milvang-Jensen\inst{2}
   \and 
   L. Kaper\inst{3}
   \and 
   D. Malesani\inst{2}
   \and 
   J.P.U. Fynbo\inst{2}
   \and 
   Z. Cano\inst{18}
   \and 
   S. Covino\inst{12}
   \and 
   H. Flores\inst{17}
   \and 
   S.Greiss\inst{4}
   \and 
   F.~Hammer\inst{17}
   \and 
   O.E. Hartoog\inst{3}
   \and 
   S. Hellmich\inst{19}
   \and 
   C. Heuser\inst{20}
   \and 
   J. Hjorth\inst{2}
   \and 
   P. Jakobsson\inst{18}
   \and 
   S. Mottola\inst{19}
   \and 
   M.~Sparre\inst{2}
   \and 
   J.~Sollerman\inst{21,22}
   \and 
   G. Tagliaferri\inst{12}
   \and 
   N.R. Tanvir\inst{11}
   \and 
   M. Vestergaard\inst{2,23}
   \and 
   R.A.M.J. Wijers\inst{3}
          }

   \institute{Instituto de Astrof\' isica de Andaluc\' ia (IAA-CSIC), Glorieta de la Astronom\' ia s/n, E-18008, Granada, Spain.\\
                     \email{adeugartepostigo@gmail.com}
         \and
             Dark Cosmology Centre, Niels Bohr Institute, Juliane Maries Vej 30, Copenhagen \O, D-2100, Denmark.
         \and
             Astronomical Institute `Anton Pannekoek', University of Amsterdam, Postbus 94249, 1090 GE Amsterdam, the Netherlands.
          \and
              Department of Physics, University of Warwick, Coventry, CV4 7AL, United Kingdom.
          \and
              Unidad Asociada Grupo Ciencia Planetarias UPV/EHU-IAA/CSIC, Departamento de F\'isica Aplicada I, E.T.S. Ingenier\'ia, Universidad del Pa\'is-Vasco UPV/EHU, Alameda de Urquijo s/n, E-48013 Bilbao, Spain.   
          \and
              Ikerbasque, Basque Foundation for Science, Alameda de Urquijo 36-5, E-48008 Bilbao, Spain.
          \and
              APC, Univ. Paris Diderot, CNRS/IN2P3, CEA/IRFU, Obs. de Paris, Sorbonne Paris Cit\'e, France.
          \and
              Pontificia Universidad Cat\'olica de Chile, Departamento de Astronom\'ia y Astrof\'isica, Casilla 306, Santiago 22, Chile.
          \and
              Millennium Center for Supernova Science, Chile.
          \and
              European Southern Observatory, Karl-Schwarzschild-Strasse 2, 85748, Garching, Germany.
          \and
              Department of Physics and Astronomy, University of Leicester, University Road, Leicester, LE1 7RH, United Kindom.
          \and
              INAF - Osservatorio Astronomico di Brera, Via E. Bianchi 46, I-23807 Merate, Italy.
          \and
              Mullard Space Science Laboratory, University College London, Holmbury St. Mary, Dorking, Surrey RH5 6NT, United Kingdom.
          \and
              ASI - Science Data Center, via Galileo Galilei, 00044, Frascati, Italy.
          \and
              INAF-Osservatorio Astronomico di Roma, Via Frascati 33, I-00040 Monte Porzio Catone, Rome, Italy.
          \and
              European Southern Observatory, Alonso de C\'{o}rdova 3107, Vitacura, Casilla 19001, Santiago 19, Chile.
          \and
              Laboratoire GEPI, Observatoire de Paris, CNRS-UMR8111, Univ. Paris Diderot, 5 place Jules Janssen, 92195 Meudon, France.
          \and
              Centre for Astrophysics and Cosmology, Science Institute, University of Iceland, Reykjavik, Iceland.
          \and
              Institute of Planetary Research, DLR, Rutherfordstrasse 2, 12489 Berlin, Germany.
          \and
              Dr. Karl Remeis-Observatory \& ECAP, Astronomical Institute, Friedrich-Alexander University, Erlangen-Nuremberg, Sternwartstr. 7, D 96049 Bamberg, Germany.
          \and
              Department of Astronomy, AlbaNova Science Center, Stockholm University, SE-106 91 Stockholm, Sweden.
          \and
              The Oskar Klein Centre, AlbaNova, SE-106 91 Stockholm, Sweden.
          \and
              Steward Observatory, University of Arizona, 933 North Cherry Avenue, Tucson, AZ 85721, United States.      
                      }

   \date{Received, ; accepted}
   
  \abstract
   {Short duration gamma-ray bursts (SGRBs) are thought to be related to the violent merger of compact objects, such as neutron stars or black holes, which makes them promising sources of gravitational waves. The detection of a `kilonova'-like signature associated to the {\it Swift}-detected GRB\,130603B has suggested that this event is the result of a compact object merger.}
   {Our knowledge on SGRB has been, until now, mostly based on the absence of supernova signatures and the analysis of the host galaxies to which they cannot always be securely associated. Further progress has been significantly hampered by the faintness and rapid fading of their optical counterparts (afterglows), which has so far precluded spectroscopy of such events. Afterglow spectroscopy is the key tool to firmly determine the distance at which the burst was produced, crucial to understand its physics, and study its local environment.}
   {Here we present the first spectra of a prototypical SGRB afterglow in which both absorption and emission features are clearly detected. Together with multiwavelength photometry we study the host and environment of GRB\,130603B.}
   {From these spectra we determine the redshift of the burst to be $z=0.3565\pm0.0002$, measure rich dynamics both in absorption and emission, and a substantial line of sight extinction of $A_V=0.86\pm0.15$ mag. The GRB was located at the edge of a disrupted arm of a moderately star forming galaxy with near-solar metallicity. Unlike for most long GRBs (LGRBs), $N_{\rm H_X}/A_V$ is consistent with the Galactic ratio, indicating that the explosion site differs from those found in LGRBs. }
   {The merger is not associated with the most star-forming region of the galaxy; however, it did occur in a dense region, implying a rapid merger or a low natal kick velocity for the compact object binary.}

   \maketitle
\section{Introduction}

Gamma-ray bursts (GRBs) are usually classified, through their gamma-ray emission, into two distinctive types: long, with typical durations of more than 2 s and soft spectra, or short, lasting less than 2 s and with harder spectra \citep{kou93}. The former class has been studied in detail for over 15 years, with optical spectroscopy of their afterglows available for over 200 of them \citep[e.g.][]{fyn09,deu12, cuc13b}. Their broad-band emission is powered by synchrotron radiation, they explode in star forming galaxies, and are accompanied by supernova emission \citep[detected in almost all cases close enough to be studied in detail;][]{hjo12}, which establishes that they are produced in the collapse of massive stars. On the other hand, short bursts have significantly fainter afterglows, which makes them much more elusive. Until now, no optical spectrum of sufficient quality to detect absorption lines has been obtained for the afterglow of a prototypical short GRB afterglow. Our observational knowledge of their properties is based on their association with the host galaxies to which they are linked \citep{fon13}, an exercise that is not always straightforward \citep{ber10}. The lack of an associated supernova and their link to a heterogeneous sample of host galaxies \citep{blo06,kan11} is consistent with an origin in a compact binary merger \citep{li98,ros03}, such as neutron stars or black holes, with a broad and uncertain range of merger times \citep{bel06}. Such events are some of the most promising sources of gravitational waves \citep{bel02}, making their study relevant in several fields of astrophysics.

Since the discovery of the first optical counterpart in 2005 \citep{hjo05,fox05,cov06}, there have been several unsuccessful attempts to obtain spectra of a short-hard GRB. In some cases, the spectra were too faint or host galaxy dominated, not allowing measurements of the line of sight of the GRB afterglow (\citealt[f.ex. GRB\,051225A,][]{sod06} or \citealt[f.ex. GRB\,070707,][]{pir08}. In other cases the afterglow spectra were obtained but the classification of the GRB was ambiguous, such as for GRB\,090426 \citep{ant09,lev10b,tho11}, or GRB\,100816A \citep{tan10,gor10}.

In this paper we present the discovery and study of the optical counterpart of GRB\,130603B, including the first absorption spectroscopy obtained for a prototypical short-hard GRB. Together with the discovery of an associated `kilonova' \citep{tan13,ber13} this burst provides us with a unique opportunity to understand the nature and conditions in which a SGRB is produced. Further spectroscopic observations of GRB\,130603B and/or its host galaxy were also reported by other groups \citep{cuc13,san13,fol13}.

The paper is structured as follows: Section 2 presents the observations and the methods used in our work, section 3 presents the results of the analysis and discusses their implications, section 4 gives our conclusions.

\section{Observations}

\subsection{Prompt emission and classification}
   
\begin{figure}[h!]
   \centering
   \includegraphics[width=\hsize]{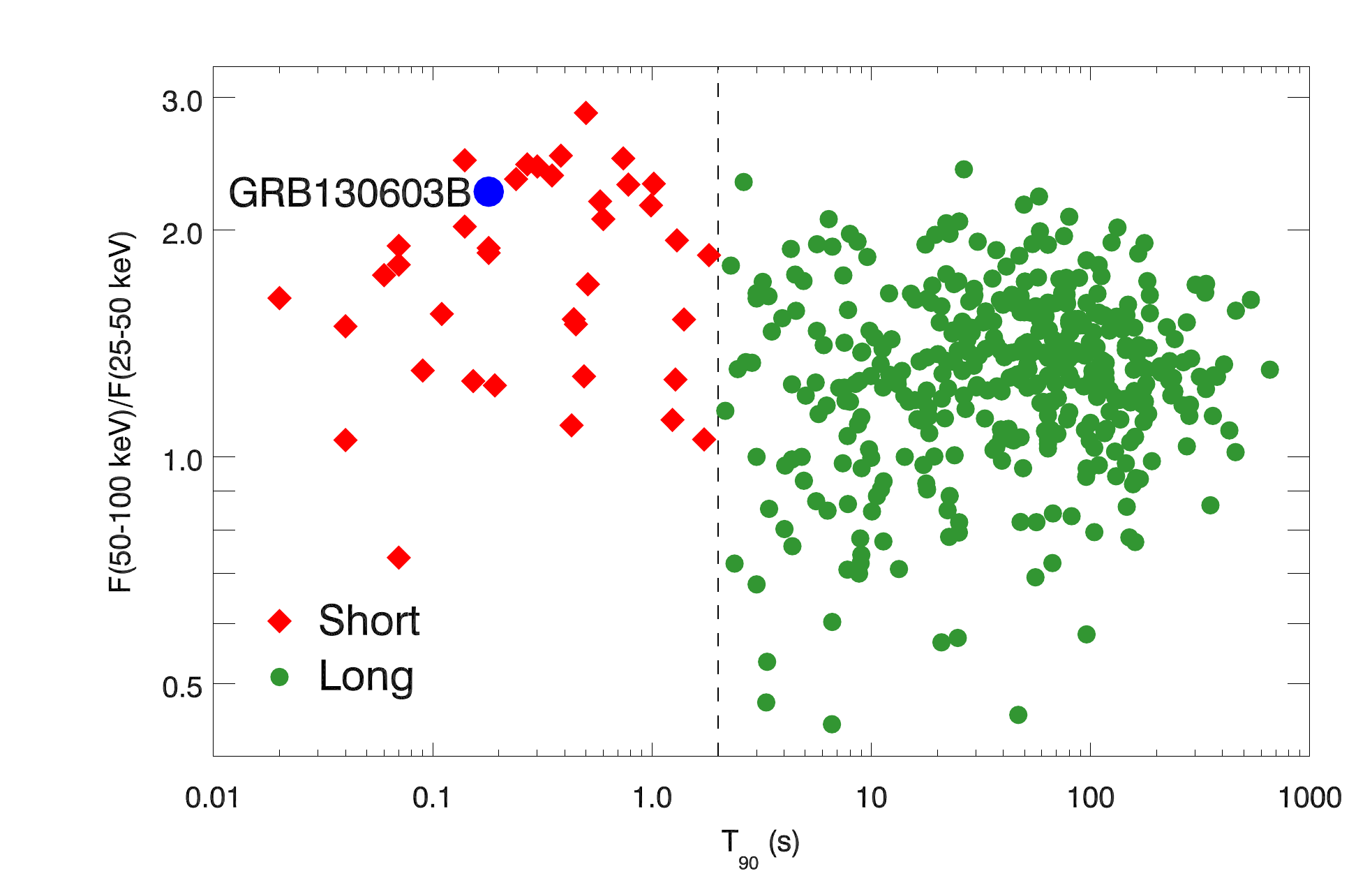}
      \caption{Hardness ratio versus $T_{90}$ duration diagram \citep[based on][]{kou93}, with the classical 2 s division, showing that GRB\,130603B belongs to the short-hard GRB family. GRB sample from {\it Swift}/BAT \citep{sak11}.
              }
         \label{Fig:hr}
   \end{figure}

GRB\,130603B triggered the Burst Alert Telescope (BAT) onboard the \textit{Swift} satellite at 15:49:14 UT on June 3 2013 \citep[$t_0$ hereafter;][]{mel13}. It had a $T_{90}$ duration of $0.18\pm0.02$ s in the 15-350 keV band and a hardness ratio (HR) $F$(50-100 keV)/$F$(25-50 keV)=$2.25\pm0.16$ \citep{bar13}, placing it in the middle of the distribution of short-hard bursts in the HR vs. $T_{90}$ diagram (see Fig.~\ref{Fig:hr}). Using the method described in \citet{hor10}, based on the location of the burst in this diagram we obtain a probability of 99.9997\% of the burst belonging to the short-hard GRB family. Observations of the event by Konus/\textit{Wind}, covering the 20 keV -- 10 MeV range, reveal an observed peak energy of $E_{{\rm peak,observed}}=660\pm100$ keV, or $E_{\rm peak,rest}=895\pm135$ keV in the rest frame of the GRB, in the typical range of short-hard bursts.  A  spectral lag analysis reveals delays of $0.6\pm0.7$ ms between the 15-25 keV and the 50-100 keV bands and $-2.5\pm0.7$ ms between the 25-50 keV and the 100-350 keV bands \citep{nor13}. These negligible or even negative lags are typical for short-hard bursts \citep{nor06}. Finally, the BAT light curve reveals no trace of extended emission at the 0.005 count det$^{-1}$ s$^{-1}$ level \citep{nor10,nor13}. Based on the above lines of evidence, there is no doubt that GRB\,130603B is a genuinely short GRB.

\subsection{Photometric observations}

Following the detection of the burst by {\it Swift} we started a follow-up campaign and discovered a relatively bright optical counterpart with the WHT \citep{lev13} and NOT \citep{deu13} telescopes. Optical and near-infrared (NIR) photometry was obtained with the 1.23m CAHA, GTC, NOT, WHT, TNG, VLT and Gemini telescopes. We also include in our study the ultraviolet and optical observation from the UVOT telescope onboard the \textit{Swift} satellite. Late observations with the Hubble Space Telescope revealed the host as a tidally disrupted late type galaxy \citep{tan13, ber13}. 

The field lies within the SDSS footprint and so our photometric calibration for $griz$ observations is obtained directly from it. Our NIR calibration is taken from 2MASS, while we utilise our own calibration based on observations of standard fields for the \textit{V}-band. 

In order to separate the contribution of the host galaxy from the afterglow light, we performed image subtraction with the public ISIS code \citep{ala00}. For clean subtractions we selected a later time (afterglow free) image from each telescope as a template and subtract this from the earlier data. Photometric calibration of these subtracted images was obtained by creating artificial stars of known magnitude in the first image, with the errors estimated from the scatter in a large number of apertures (of radius approximately equal to the seeing) placed within the subtracted image. The placement of artificial stars close to the limiting magnitude within the image confirms that these can be recovered, and so the given limiting magnitudes are appropriate. However, we do note that the limiting magnitudes are based on the scatter in photometric apertures placed on the sky, not on the relatively bright regions of the host directly underlying the GRB. 
The log of optical/NIR photometry of the afterglow is given in Table~\ref{table:photlog}.

The analysis performed on the \textit{Swift}/UVOT data is significantly different from the ground based photometry. Due to this, we display the results separately from the optical/NIR data. The data tables and details of the UVOT analysis are provided in the Appendix. We used the stacked UVOT data (Table~\ref{table:uvot2}), where the last epoch was subtracted from the rest, assuming that this last one had only emission from the host galaxy. These data are important, as they cover the early phase of the optical light curve and allow us to identify a flattening in the optical light curve which significantly differs from the X-ray emission. 

\begin{table}[htdp]
\caption{Photometric observations of the GRB 130603B afterglow (host subtracted). Note that the VLT/FORS2 $V$-band photometry is measured in a small (0.7$^{\prime\prime}$) aperture, but is not host subtracted. The errors
given are statistical only and do not account for systematics between slightly different filter systems. The details
of the subtraction also introduces small differences to the recovered flux, especially for sources sitting on moderately 
bright extended regions of their host galaxies. To account for this we estimate the additional variance on
artificial stars inserted into the images to be $\sim 0.05$ mag. }
\label{table:photlog}
{\scriptsize 
\centering                          
\begin{tabular}{llllll}
\hline\hline                 
$t-t_0$ (days) & Telescope/Instrument & Band & Exposure (s)  & AB Magnitude  \\
\hline
  0.244	& NOT/MOSCA		& $r$ 	& $5 \times 360$ 	& 21.15 $\pm$ 0.02 	\\
  0.254 	& WHT/ACAM 		& $z$ 	& $3 \times 300$ 	& 20.39 $\pm$ 0.06 	\\
  0.274 	& WHT/ACAM 		& $i$ 	& $3 \times 300$ 	& 20.86 $\pm$ 0.06 	\\ 
  0.281	& 1.23m CAHA		& $V$	& $6 \times 300$	& 21.60 $\pm$ 0.10 \\
  0.290 	& GTC/OSIRIS 		& $r$ 	& 30				& 21.30 $\pm$ 0.02 	\\
  0.294 	& WHT/ACAM 		& $g$ 	& $3 \times 300$ 	& 21.90 $\pm$ 0.06 	\\
  0.329 	& VLT/FORS2 		& $V$ 	& $60$ 			& 21.47 $\pm$ 0.02	\\
  0.599 	& Gemini-N/GMOS-N& $z$ 	& $5 \times 100$ 	& 21.94 $\pm$ 0.03 	\\
  0.603 	& UKIRT 			& $K$  	& $70 \times 10$ 	& 21.06 $\pm$ 0.11 	\\
  0.607 	& Gemini-N/GMOS-N & $i$ 	& $5 \times 100$   	& 22.34 $\pm$ 0.03 	\\
  0.614 	& UKIRT 			& $J$ 	& $70 \times 10$ 	& 21.48 $\pm$ 0.14 	\\
  0.615 	& Gemini-N/GMOS-N 	& $r$ 	& $5 \times 100$  	& 22.83 $\pm$ 0.03 	\\
  0.623 	& Gemini-N/GMOS-N 	& $g$ 	& $5 \times 100$  	& 23.47 $\pm$ 0.04 	\\
  1.59 	&  Gemini-N/GMOS-N 	& $g$ 	& $5 \times 120$ 	& $>$25.7 		\\
  1.60 	&  Gemini-N/GMOS-N 	& $r$ 	& $5 \times 120$ 	& 25.6$\pm$0.3  		\\
  1.61	& UKIRT			& $J$	& $140\times10$	& $>$22.5			\\
  1.61 	&  Gemini-N/GMOS-N 	& $i$ 	& $5 \times 120$ 	& $>$24.7  		\\
  1.62 	& Gemini-N/GMOS-N 	& $z$ 	& $5 \times 120$ 	& $>$23.9 		\\
  2.32 	& VLT/HAWK-I  			& $J$ 	& $22 \times 60$ 	& $>$24.3  		\\
  3.26 	& GTC/OSIRIS 			& $r$ 	& $3 \times 200$ 	& $>$25.1  		\\
  4.26 	& GTC/OSIRIS 			& $r$  	& $3 \times 200$ 	& $>$25.5  		\\
  7.30	& VLT/HAWK-I			& $J$	& $22\times$60	& $>$23.6			\\
  8.23 	& TNG/DOLoRes	& $r$	& $4\times300$ 	&  $>$23.75			\\
  8.25	& TNG/DOLoRes	& $i$	& $4\times300$   	&  $>$24.61			\\
  21.26	& TNG/DOLoRes	& $r$	& $8\times300$    	&  $>$23.47			\\
  21.29	& TNG/DOLoRes	& $i$	& $8\times300$    	&  $>$23.98			\\
\hline                                   
\end{tabular}
}
\end{table}

The X-Ray Telescope (XRT) onboard {\it Swift} began observing the GRB field on 2013 Jun 03 at 15:49:13.945 UT. We used photon counting (PC) mode data for the X-ray spectral analysis. We used the reductions provided by the {\it Swift} Burst Analyser \citep{eva10}, transforming them to 2 keV. 

The X-ray light curve is characterised by a slow decay during the first 0.1 days, followed by a gradual steepening. The late optical data match reasonably well the steep late decay of the X-ray light curve, which makes the afterglow undetectable after one day even for large telescopes. There is a clear break in the optical light curve at $\sim0.25$ days before which the evolution strongly differs from the X-ray one, with the optical being much flatter than the X-rays or even consistent with a brightening until 0.2 days. 

\subsection{Spectroscopy}

Optical spectroscopy with OSIRIS, at the Gran Telescopio Canarias (GTC), starting 7 hours after the GRB onset, revealed both absorption and emission features at a common redshift of 0.3565$\pm$0.0002 \citep{tho13}. Three more spectra were obtained in the hours that followed with FORS2 and X-shooter \citep{xu13} at the Very Large Telescope and ACAM at the William Herschel Telescope. 

\begin{table}[ht!]
\caption{Log of spectroscopic observations.} 
\label{table:speclog}      
{\scriptsize \begin{center}                        
\begin{tabular}{c c c c c c}       
\hline\hline               
Mid $t-t_0$      	& Telescope/			&  Range 		& Resolution		& Exposure  		\\
(hr)			& Instrument    			&  ({\AA})         		&                    		&     (s)                      	\\
\hline                        
7.410		& GTC/OSIRIS 			& 3650--7800	& 1000			& $3\times900$ 	\\
8.220		& WHT/ACAM 				& 3500--9500	& 530			& $1\times900$ 	\\
8.404		& VLT/FORS2			& 3600--9150	& 590			& $3\times600$	\\
8.555		& VLT/X-shooter		& 3000--24800	& 5100/8800/5300	& $4\times600$ 	\\
126.04		& GTC/OSIRIS			& 3650--7800	& 1000			& $3\times1200$ 	\\
\hline                                   
\end{tabular}
\end{center}}
\end{table}

\begin{figure}[ht!]
   \centering
   \includegraphics[width=\hsize]{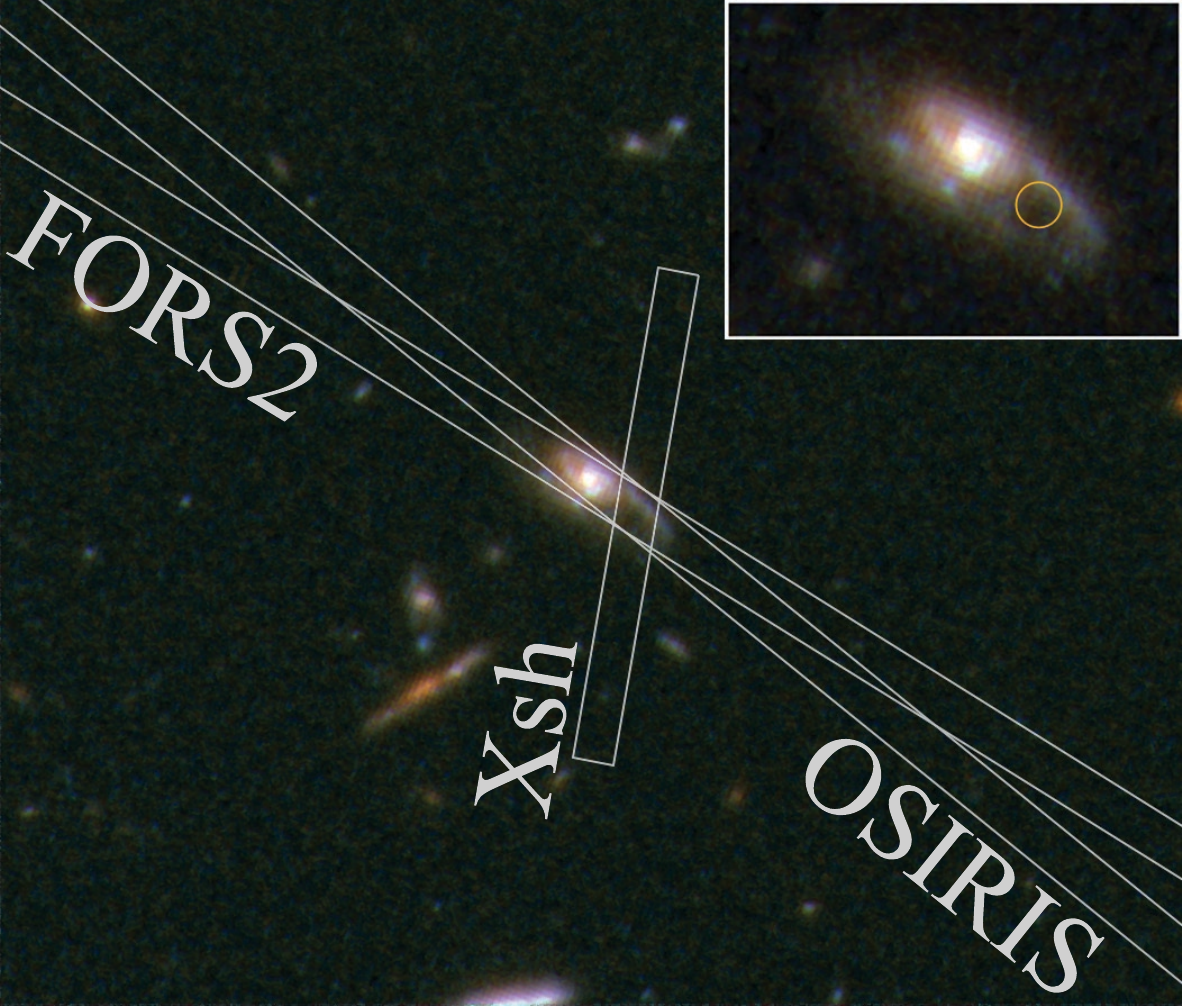}
      \caption{Colour composite based on the HST observations \citep{tan13} in which the positions of the slits of the OSIRIS/GTC, X-shooter/VLT and FORS/VLT have been marked. North is to the top and East to the left, the field of view is $26^{\prime\prime}\times22^{\prime\prime}$. The inset shows a blow-up of the central $6^{\prime\prime}\times4^{\prime\prime}$, with the position of the afterglow marked with a circle.
              }
         \label{fig:finder}
   \end{figure}

The OSIRIS spectrum was obtained with the slit placed along the major axis of the host galaxy and covering the afterglow under moderate seeing conditions ($\sim1^{\prime\prime}$). The second spectrum was taken 5 days after the burst with identical settings and similar seeing conditions. 
The FORS2 spectrum was aligned in a similar angle but had better seeing ($\sim0.6^{\prime\prime}$), allowing us to extract the afterglow spectrum with minimum contamination from the host. For the X-shooter spectrum the slit was aligned almost perpendicular to the host axis and it includes only a small host contribution from the neighbourhood of the GRB. A diagram with the slit position of the most relevant observations is shown in Fig.~\ref{fig:finder}.

\section{Results and discussion}

\subsection{Spectral features along the line-of-sight}

\begin{figure}[h!]
   \centering
   \includegraphics[width=\hsize]{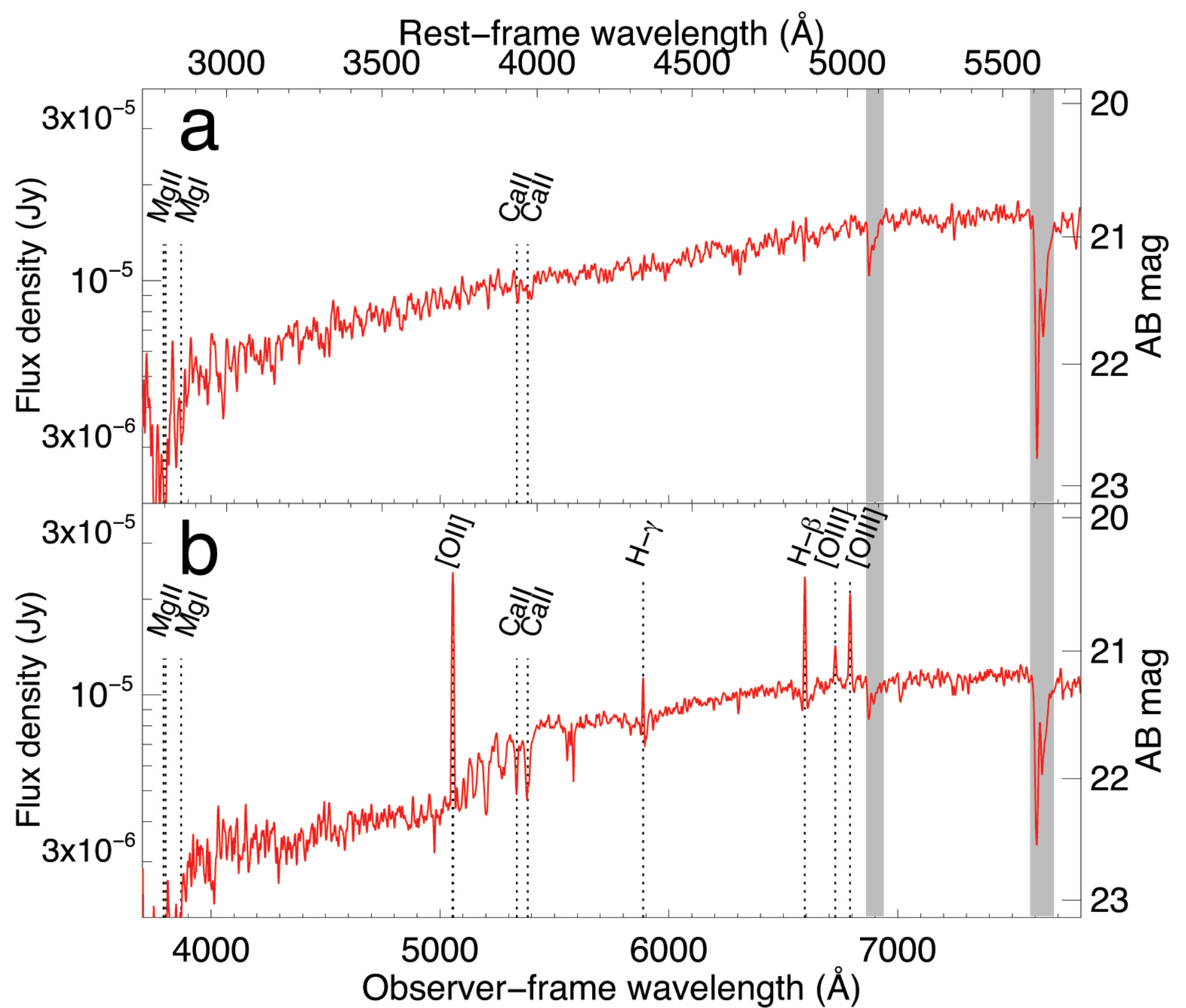}
      \caption{(a) Spectrum of the afterglow of GRB\,130603B obtained using the OSIRIS instrument on the 10.4 m GTC telescope 7.4 hr after the burst onset. The contribution of the host galaxy was subtracted using a second spectrum, that was obtained 5 days later when the afterglow had faded away. Panel (b) shows this second spectrum of the host galaxy. Dotted vertical lines indicate the location of some of the absorption and emission features. Grey vertical bands mark the position of strong telluric features.
              }
         \label{Fig:specs}
   \end{figure}

The first spectrum of GRB\,130603B was obtained by OSIRIS with a seeing of ~1.0$^{\prime\prime}$. This implied that there was some degree of blending between the GRB afterglow and the host galaxy. In order to obtain a clean spectrum of the afterglow, with no contamination from the host galaxy, we used the second spectrum obtained 5 days later, when the afterglow contribution was negligible, and subtracted it from the early data. The result was a GRB afterglow spectrum, clean of emission lines, but with absorption features due to \ion{Mg}{ii} and \ion{Ca}{ii} (see Fig.~\ref{Fig:specs}). The lack of residuals in correspondence of the bright emission lines highlights that our procedure is reliable, and confirms that the observed absorption lines were indeed features of the afterglow spectrum.

The spectra obtained at the VLT one hour later had better atmospheric conditions. This allowed us to separate well the emission of the host galaxy core from the afterglow. Using the host galaxy spectrum of the GTC, we estimate that the contribution of the host galaxy in these spectra is less than 30\% in the $r$-band (where the contribution is strongest) and less than 20\% bluewards of the $g$-band, where the Mg lines are located. In these spectra we can measure the same absorption features that are seen in the clean GTC spectrum with consistent equivalent widths. In addition, absorptions from \ion{Mg}{i} and \ion{Na}{i} are also detected at the same redshift. Table~\ref{table:EW} displays the equivalent width values measured for each absorption line in the observer frame.

   \begin{figure}[ht!]
   \centering
   \includegraphics[width=8.1cm]{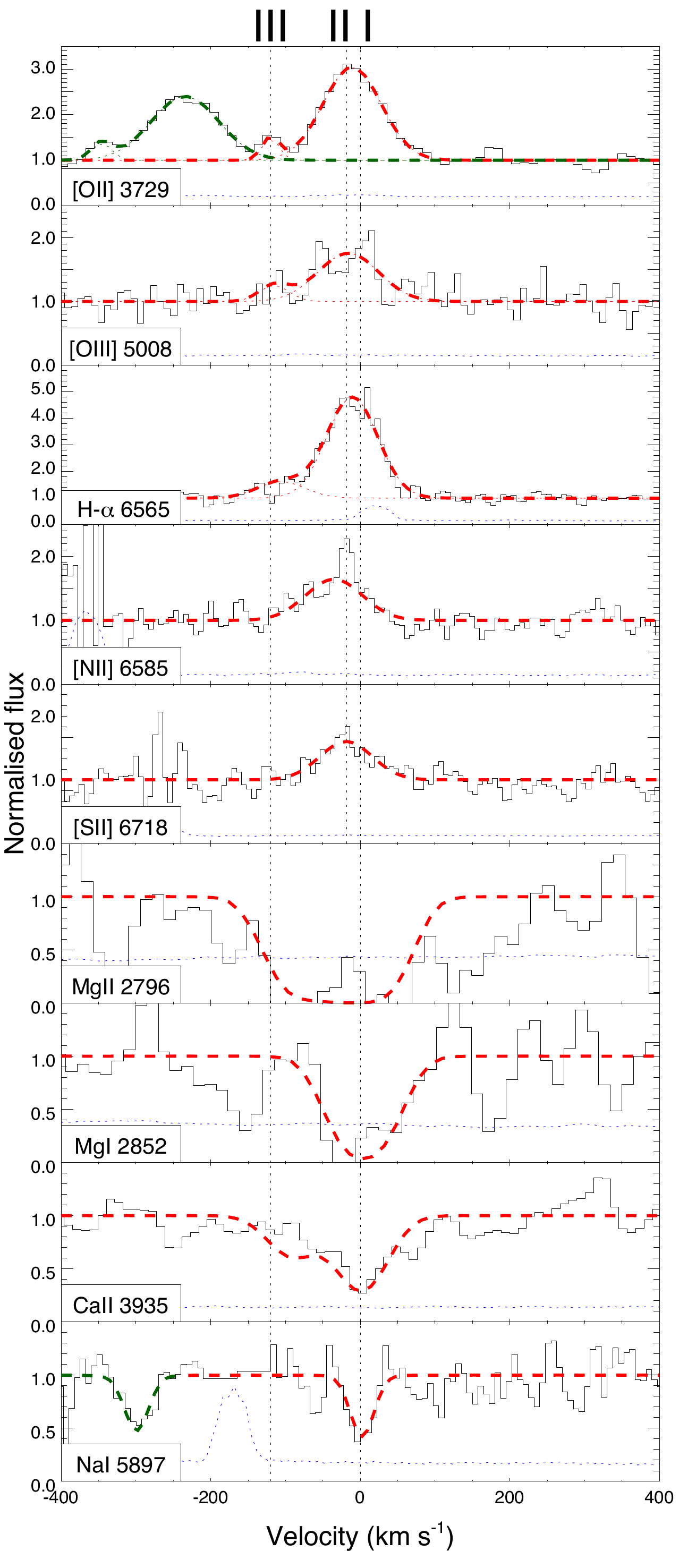}
      \caption{Selection of some of the emission and absorption features in the line of sight of GRB\,130603B as observed in the X-shooter afterglow spectrum obtained 8.6 hr after the burst onset. Black lines correspond to the normalised spectrum in velocity space, centred at each of the lines, assuming a redshift of 0.3565. Blue dotted lines are the corresponding error spectra. Red dashed lines are Gaussian fits to the emission features and Voigt profile fits to the absorption features. Green dashed lines are used when a further line of the same species, which has also been fitted, lies within the plotted range. Vertical dotted lines mark the position of the different velocity components, where I is the main absorption feature, II is the main emission feature at $-18$ km s$^{-1}$ and III is an additional feature at $-120$ km s$^{-1}$.
              }
         \label{Fig:lines}
   \end{figure}

\begin{table}[ht!]
\caption{Equivalent widths of the detected species (observer frame) in our spectra.}             
\label{table:EW}      
\begin{center}                        
\begin{tabular}{c c c c}       

\hline\hline                        
Feature						&	X-shooter			&	GTC				&	FORS		\\
							&	EW ({\AA})		&	EW ({\AA})		&	EW ({\AA})			\\
\hline                        
\ion{Mg}{ii} $\lambda$2796		&	$3.8\pm0.6$		&	$4.6\pm1.6$		&	$6.4\pm0.8$		\\
\ion{Mg}{ii} $\lambda$2803		&	$2.3\pm0.5$		&	$3.1\pm1.3$		&	(blended)			\\
\ion{Mg}{i} $\lambda$2853		&	$1.36\pm0.20$		&	---				&	$1.5\pm0.4$			\\
\ion{Ca}{ii} $\lambda$3935 		&	$2.52\pm0.18$		&	$2.1\pm0.4$		&	$2.3\pm0.8$		\\
\ion{Ca}{ii} $\lambda$3970        	&	$1.09\pm0.12$		&	$1.3\pm0.4$		&	$1.4\pm0.6$		\\
\ion{Na}{i} $\lambda$5892       	 	&	$0.53\pm0.09$		&	---				&	$1.9\pm0.6$			\\
\ion{Na}{i} $\lambda$5898       	 	&	$0.59\pm0.08$		&	---				&	(blended)			\\

\hline                                   
\end{tabular}
\end{center}
\end{table}

The X-shooter spectrum had enough spectral resolution to resolve the features, both in emission and in absorption (see Fig.~\ref{Fig:lines}). The emissions have at least two components at the line of sight of the GRB, the main one is slightly blueshifted (component II, in Fig.~\ref{Fig:lines}, at $\sim-18$ km s$^{-1}$) from the the absorption features (which we identify as component I) and has a full width half maximum (FWHM) of $92\pm9$ km s$^{-1}$. A minor emission component (III) is seen at a velocity of $\sim -120$ km s$^{-1}$.

The absorption features are broad, but a Voigt profile fitting does not provide significant constraints to the column densities of the atomic species that create them due to the low signal-to-noise ratio. The absorbers also seem to show rich dynamics, with line widths corresponding to velocities of 100--180 km s$^{-1}$. Although it is hard to separate the broadening due to saturation from the one due to dynamics, there seems to be some correlation between the absorption and emission components in the line of sight of the GRB, pointing to a dynamical origin. We notice that, while the Mg and Ca lines show a broad profile, the \ion{Na}{i} remains unresolved at a resolution of $\sim30$ km s$^{-1}$.

To study the absorption properties, we measure the equivalent widths of each of the features in the different spectra, as displayed in Table~\ref{table:EW}. These absorption features can be compared to the sample of absorption lines in the line of sight of long GRB spectra \citep{fyn09,deu12}. A way of doing this is by using the line strength parameter \citep[LSP;][]{deu12}. With a LSP = 0.20$\pm0.13$, GRB\,130603B has absorption features that are just slightly stronger than the average long GRB spectrum, indicating a dense environment along the line of sight of the GRB. Similar results are obtained if the comparison with the sample is done only the Mg lines (LSP$_{\rm Mg}=0.25\pm0.08$) or only with the Ca lines (LSP$_{\rm Ca}=0.16\pm0.08$).

\subsection{SED fit and extinction analysis}

The shape of the continuum in the optical afterglow spectrum presents a strong curvature due to extinction in the line of sight to the GRB. Together with the spectrum obtained in X-rays by {\it Swift}/XRT at a similar epoch, the SED can be well fitted \citep[following the method described by][]{zaf11} by a broken power law with spectral indices of $\beta_{\rm O}=-0.65\pm0.09$ and $\beta_{\rm X}=-1.15\pm0.11$ (where the flux density relates with frequency as $F_{\nu}\propto \nu^{\beta}$) in the optical and X-rays, respectively (more details on the fitting method and analysis are given in the Appendix). These values are consistent with a single synchrotron emission component (optically thin, with the cooling break between optical and X-rays and an electron energy index of $p=2.3$) and line of sight extinction of $A_V=0.86\pm0.15$ mag with an extinction law consistent with that of the Small Magellanic Cloud. This relatively high extinction is in agreement with the detection of \ion{Na}{i} absorption, which is known to be correlated with dust extinction \citep{poz12}, although with a large scatter \citep{phi13}.

\begin{figure}[ht!]
   \centering
   \includegraphics[width=\hsize]{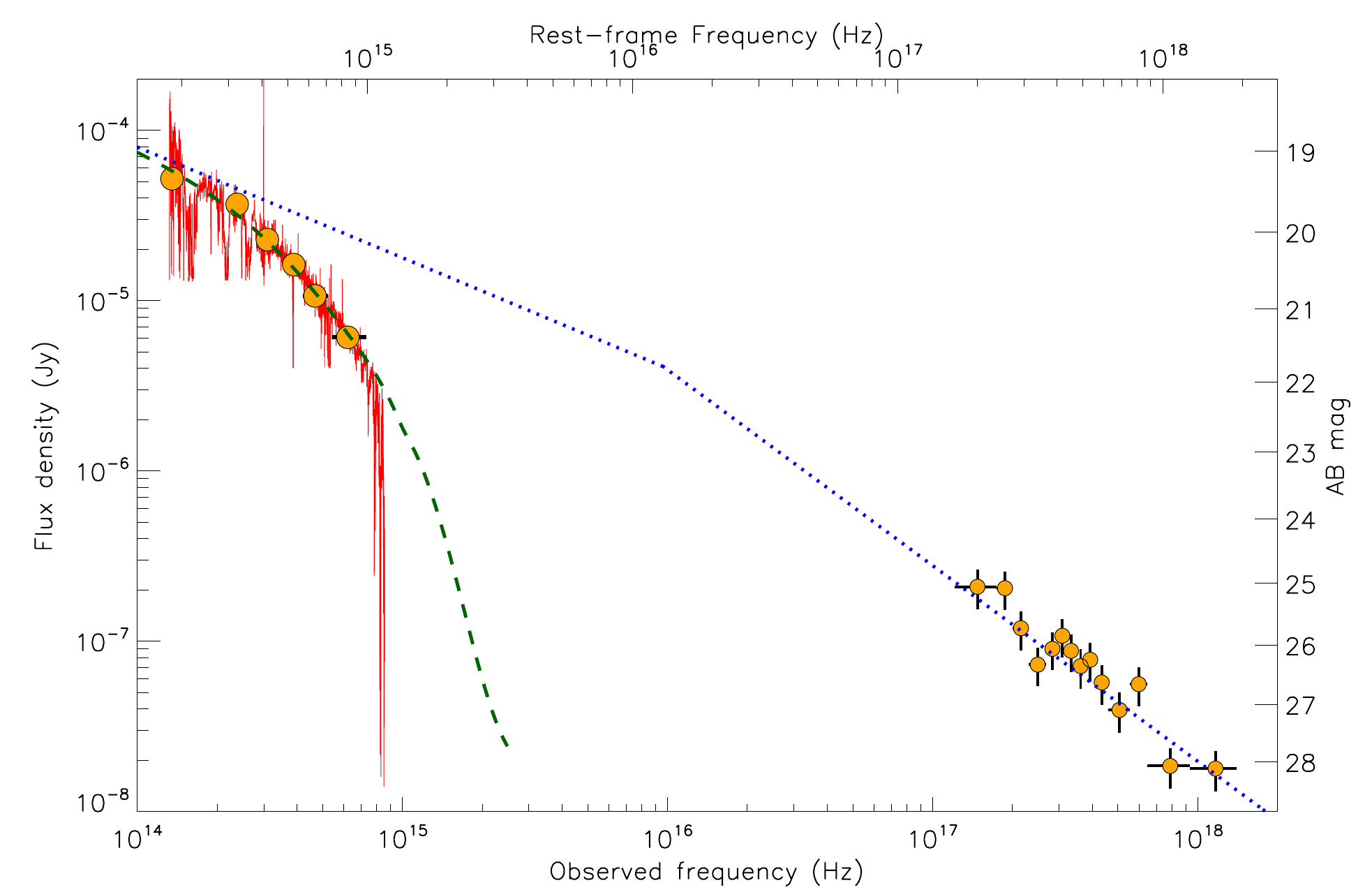}
      \caption{NIR to X-ray spectral energy distribution of GRB\,130603B 8.5 hr after the burst onset. The spectrum (in red) can be fitted by a broken power law (blue dotted line) with dust extinction in the line of sight (green dashed line). Orange dots are the photometric observations scaled to the time of the fitted spectra. X-ray data are absorption corrected.
              }
         \label{Fig:SED}
   \end{figure}

Long GRBs show a surprisingly high absorption in X-rays due to metals in the line of sight ($N_{\rm H_X}$), which is about one order of magnitude larger \citep[e.g.][]{cam10} than one would expect from the extinction caused by dust seen in the optical. Other objects in the Universe show a fairly uniform $N_{\rm H_X}$ to dust ratio \citep{che13}.  This discrepancy might be due to the close proximity of high density gas close to the burst causing the X-ray absorption \citep{wat13}.  Our spectra of GRB\,130603B show extinction and a neutral metal column ($N_{\rm H_X}/A_V=(1.56\pm0.20)\times10^{21} $ cm$^{-2}$ mag$^{-1}$) that are consistent with the Galactic (i.e. normal) ratio, indicating that the explosion site for this  short GRB was not like the star-forming regions in which long GRBs explode. The average value of $N_{\rm H_X}/A_V$, together with the significant dust extinction, indicate that the binary is outside of its birth cloud but still in a dense region of the galaxy, suggesting a  short, but not instantaneous, merger process.
   
\subsection{Light curve analysis}

\begin{figure}[ht!]
   \centering
   \includegraphics[width=\hsize]{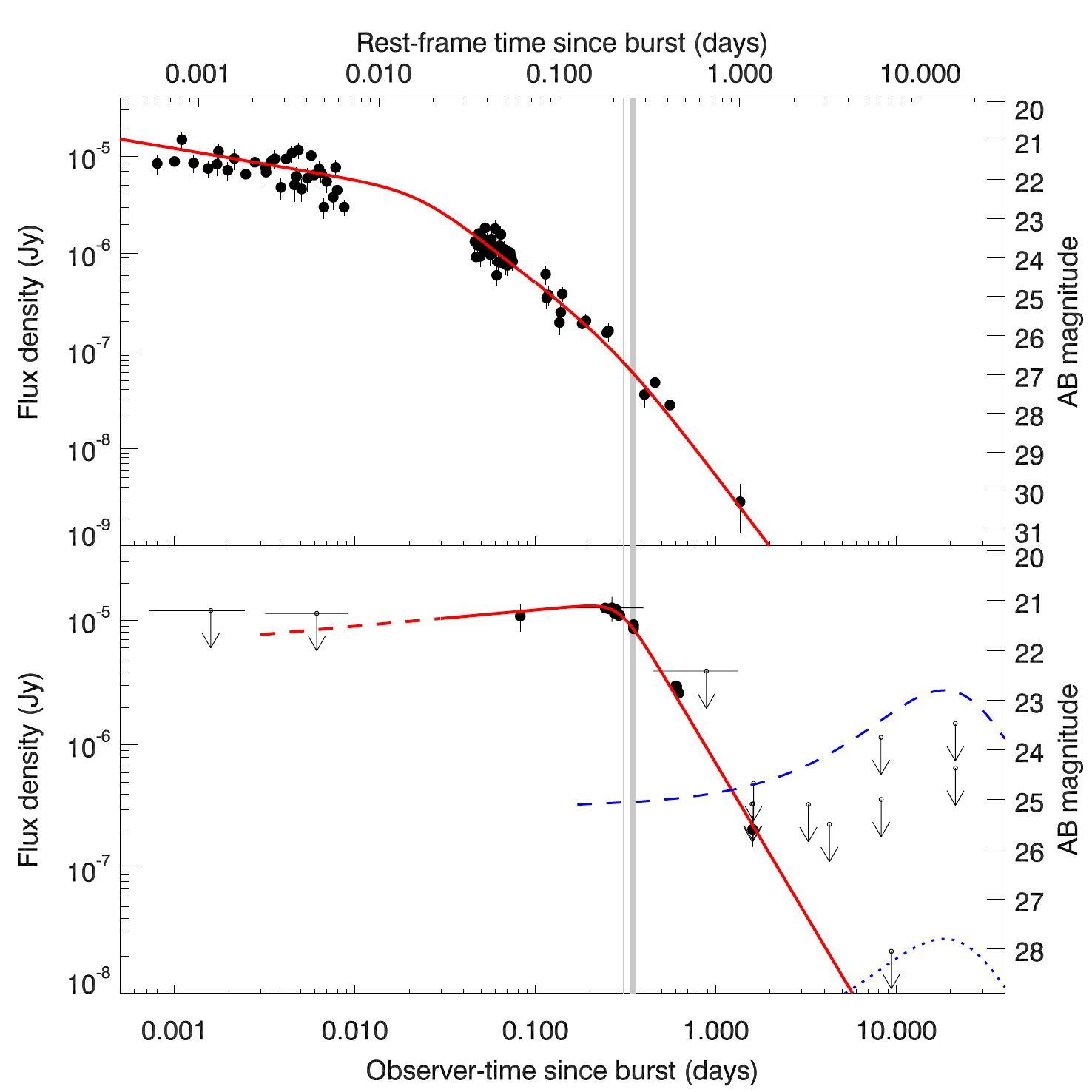}
      \caption{Light curves of GRB\,130603B at X-ray (top) and optical (bottom) wavelengths. Detections are indicated with dots and upper limits (3-$\sigma$) with arrows. The optical photometry of the different bands has been scaled to the $r$-band. The vertical lines indicate the times when spectra were obtained. We include optical data from the literature \citep{cuc13, tan13}. The red lines are fits of the light curves with broken power laws, which show that, while the evolution of X-rays and optical are very different until 0.3 days, their evolution is consistent after that time. The dashed blue line is the expected \textit{r}-band light curve of a supernova like SN~1998bw, the most common template for long GRBs, after including an extinction of $A_V=0.86$ mag. The most constraining limits indicate that any supernova contribution would be at least 100 times dimmer than SN~1998bw in the $r$-band, once corrected of extinction (blue dotted line). However, see \citet{tan13} and \citet{ber13} for the detection of a `kilonova' component at infrared wavelengths. 
              }
         \label{Fig:lc}
   \end{figure}

The X-ray light curve (see Fig.~\ref{Fig:lc}) is characterised by a slow decay during the first 0.1 days, followed by a gradual steepening. The early optical light curve differs strongly from the X-ray slope until 0.3 days, before which it is much flatter or even brightening. After the 0.3 day break the optical evolution is consistent with the late decay of the X-ray light curve making the afterglow undetectable after two days even for the largest telescopes. The late evolution could be fitted with a fireball peaking at $\sim0.2$ days with a jet break at 0.3 days and an electron energy distribution index of $p\sim2.3$.  However, the X-ray emission would only be consistent with this if there is an additional component prior to 0.1 days. The fireball is also peaking at unusually late timescales which implies that the initial Lorentz factor was very low. 

In an alternative and more likely interpretation, the merger of two neutron stars may instead form a magnetar with a millisecond spin period \citep{dai98} which can power a plateau phase as it spins down \citep{zha01}. Using this model, the X-ray light curve can be fitted with a newly-formed stable magnetar with a magnetic field of $(8.59 \pm 0.19) \left( \frac{\epsilon}{1-\cos\theta} \right) ^{0.5} \times 10^{15}$~G and a spin period of $8.44^{+0.41}_{-0.39} \left( \frac{\epsilon}{1-\cos\theta} \right) ^{0.5}$ms, where $\epsilon$ is efficiency for the conversion of rotational energy into X-rays and $\theta$ is the beaming angle of the magnetar emission, using the method described in \citet{row13}. \cite{fon13b} have recently used multi wavelength observations to also postulate a model with a magnetar contribution to the emission of GRB 130603B. In particular they identify a late time X-ray excess, using {\it XMM} observations, that is consistent with the magnetar fit we obtain. Further modelling of this burst as a magnetar has been presented by \citet{met13} and \citet{fan13}. More details of the comparison of the fireball and magnetar models are given in the Appendix.

\subsection{Host galaxy}

The host galaxy of GRB 130603B is a perturbed spiral galaxy as seen in high-resolution HST imaging \citep{tan13} due to interaction with another galaxy. The GRB is located in the outskirts of the galaxy in a tidally disrupted arm, at $5.4\pm0.3$ kpc from the brightest point of the host, consistent with the distribution of projected physical offsets between short GRBs and their hosts \citep{fon10,fon13c}. 

\begin{figure}[ht!]
   \centering
   \includegraphics[width=\hsize]{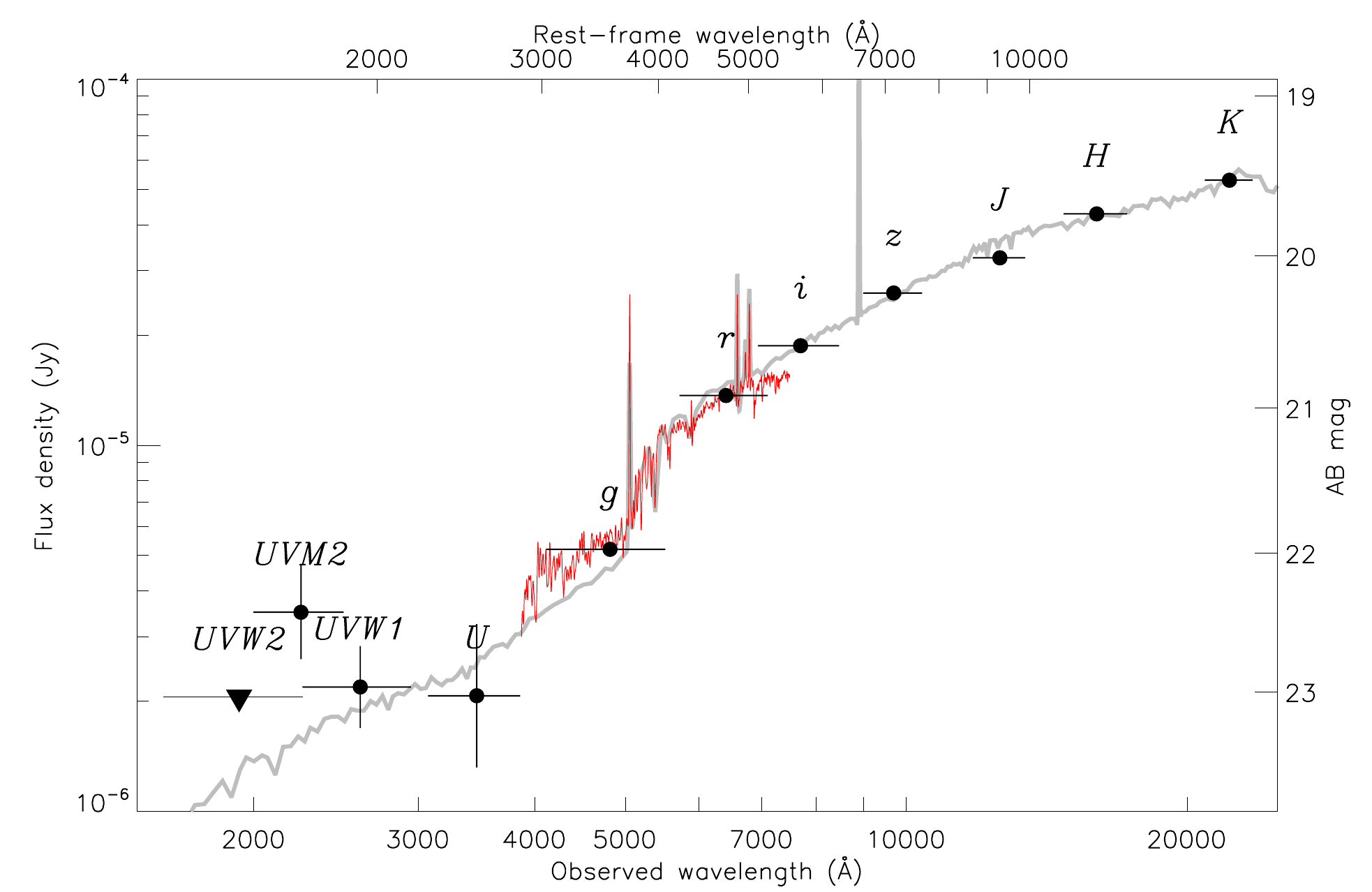}
      \caption{Ultraviolet to near infrared photometry of the host galaxy of GRB\,130603B. Dots are detections, whereas triangles are 3-$\sigma$ upper limits. Over plotted in red is the spectrum of the host galaxy obtained with OSIRIS/GTC 5 days after the burst onset, when the afterglow contribution was negligible. The grey line is the best fit to a set of galaxy templates, as described in the text.
              }
         \label{Fig:hostsed}
   \end{figure}

We obtained photometry of the host galaxy from our late time optical and IR imaging, when the afterglow was sufficiently faint not to contribute significantly. These magnitudes are generally in agreement with those in the SDSS, although typically with significantly smaller errors. From this photometry we can create the spectral energy distribution of the host from UV to NIR, as shown in Table~\ref{table:hostsed} and Fig.~\ref{Fig:hostsed}. This photometry was used to run a photometric fit to a set of host galaxy templates using LePhare \citep[version 2.2,][]{arn99,ilb06}. Templates were based on the models from \citet{bru03} created following the same procedure as in \citet{kru11}. The SED shape is best reproduced by a galaxy template with an age of $(1.0_{-0.6}^{+3.1})\times10^9$ yr, a stellar mass of log($M_*/M_\odot$)$=(9.8\pm0.2)$ (similar to the one obtained by \citealt{cuc13}, although see Sect. 3.4.2. for the mass derived from stellar population fitting), an extinction of $E(B-V)=0.3$ mag and a star formation rate of SFR = $5.6_{-3.1}^{+7.2} M_\odot$ yr$^{-1}$ (consistent with the value derived from emission line analysis, see below).

\begin{table}[ht]
\caption{Host galaxy photometry.}
\label{table:hostsed}
\centering
\smallskip
\begin{tabular}{cc}
\hline \hline 
Band		&	Magnitude     \\
\hline
\textit{uvw2}	& $>23.12$ \\
\textit{uvm2}	& $22.66^{+0.32}_{-0.25}$			\\
\textit{uvw1}	& $23.05^{+0.28}_{-0.22}$			\\
\textit{U}		& $23.11^{+0.49}_{-0.34}$			\\
\textit{g}		& $22.11\pm0.06$			\\
\textit{r}		& $21.06\pm0.06$			\\
\textit{i}		& $20.72\pm0.06$			\\
\textit{z}		& $20.36\pm0.06$			\\
\textit{J}		& $20.12\pm0.07$			\\
\textit{H}		& $19.82\pm0.06$			\\
\textit{K}		& $19.59\pm0.07$			\\
\hline
\end{tabular}
\end{table}

\subsubsection{Spatially resolved information}

The host galaxy was covered by two different slit positions along the major axis (see Sect. 2.3 and Fig.~\ref{fig:finder}). One long slit spectrum was taken with FORS at a position angle of 58$\degree$ when the afterglow was still present. Two spectra were taken with GTC/OSIRIS with a similar slit position (position angle of 51$\degree$), the first one with the afterglow present, the second one 5 days after the GRB when the afterglow had faded away.  Furthermore we have X-shooter spectra only covering the afterglow position and with the afterglow dominating the continuum. 

In the early FORS spectrum the seeing was good enough to easily separate the afterglow trace from the rest of the galaxy. We extracted separate spectra from the core and the disk of the galaxy at the opposite side of the GRB using the afterglow trace as reference. The traces have a size of 3 pixels corresponding to 0.75$''$ or 3.6\,kpc. A similar process was done using the second epoch of OSIRIS/GTC long slit observations. Here we extracted 3 different traces using the trace from the core as reference at the GRB position, the core and the opposite side of the galaxy. The pixel size is basically identical to the FORS spectrograph and the traces equally cover 3 pixels or $\sim$ 3.6\,kpc.

The parts of the galaxy extracted from the late GTC spectrum and the regions not affected by the afterglow in the FORS spectra were used to perform stellar population fitting to determine the star-formation history (see Sect. 4.3.2). The emission lines in the FORS and GTC spectra were furthermore used to analyse the properties of the ISM (see Sect. 3.4.3). As the X-shooter spectrum is afterglow dominated, its continuum cannot be used to model the host galaxy, although the emission lines can be used to determine the gas properties.

   \begin{figure*}[ht!]
   \centering
   \includegraphics[width=15cm]{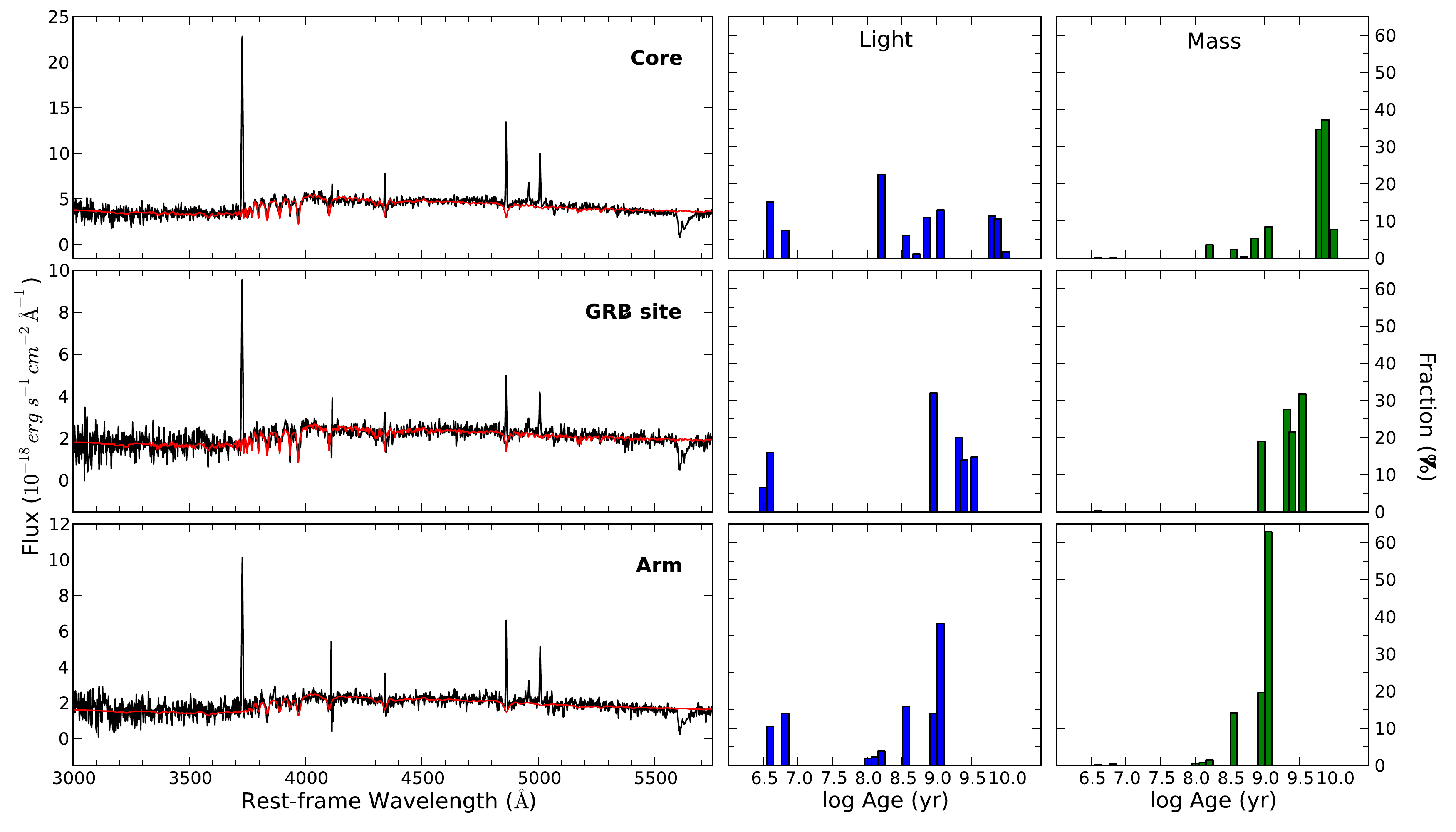}
      \caption{Stellar population fitting of spectra from three different parts of the host galaxy (using the late GTC data): Top panel: galaxy core, middle panel: GRB site, bottom panel: opposite side of the galaxy. The left panels show the original spectrum and the best fit SP model over plotted (in red), the right panels show the light and the mass fraction of populations with different ages.               }
         \label{Fig:SP}
   \end{figure*}
   
\subsubsection{Stellar population fitting}

We used the spectral synthesis code STARLIGHT\footnote{The STARLIGHT project is supported by the Brazilian agencies CNPq, CAPES and FAPESP and by the France-Brazil CAPES/COFECUB programme.} \citep{cid04, cid05}. STARLIGHT fits an observed continuum spectral energy distribution using a combination of the synthesis spectra of different single stellar populations (SSPs). We chose for our analysis the SSP spectra from \citet{bru03}, based on the STELIB library of \citet{leb03}, Padova 1994 evolutionary tracks and a Chabrier \citep{cha03} IMF between 0.1 and 100 $M_\odot$. We used a base containing 4 metallicities ($Z$ = 0.004, 0.008, 0.02 and 0.05) and 40 ages for each metallicity. The age span from 1 Myr up to 14 Gyr. 

The STARLIGHT code solves simultaneously for the ages and relative contributions of the different SSPs and the average reddening, providing the complete star formation history (SFH) extracted from the spectra. The reddening law from \citet{car89} has been used. Prior to the fitting procedure, the spectra were shifted to the rest frame and re-sampled to a wavelength interval of 1 {\AA} in the entire wavelength range. Bad pixels and emission lines were excluded from the final fits.

Figure \ref{Fig:SP} shows the SP fitting for three regions in the late GTC spectra (without afterglow contamination). The GRB site has a large fraction of an older population ($>$ 1 Gyr) but also a small, rather young population. The outer part of the galaxy opposite to the GRB regions has a similar age distribution, with an older population of somewhat lower age, around 100 Myr -- 1 Gyr. The SFH of the core is more complex and shows populations of a wide range of ages. 

The stellar population fit also returns the stellar mass of each of the regions and of the complete host, as shown in Table~\ref{table:EML-res}. The mass of the host galaxy derived this way is smaller than the one derived from the fit of the photometric SED. This is probably because the stellar population analysis fits together a range of SSPs, whereas the photometric fit was obtained assuming only one SSP. Due to this, we will assume that the mass obtained by stellar population fitting is the most reliable.

\subsubsection{Emission line analysis}

We measured the fluxes of the emission lines in the individual spectra by fitting a Gaussian profile to the lines. The X-shooter spectrum spans a wide wavelength range, including all strong emission lines. The grism used for the FORS spectrum covers all the strong optical emission lines from [\ion{O}{ii}]$\lambda$3727 to [\ion{S}{ii}]$\lambda$6733 while the GTC spectrum is missing H$\alpha$ and redder lines. The measured fluxes (not corrected for extinction) are listed in Table~\ref{table:EML}. We note that the regions covered by the different spectra are not exactly the same, due to different position angles and slit widths, which results in line fluxes that can significantly differ. Using these values we determined the properties using common strong line calibrators, depending on the lines available.  The properties of the different regions are shown in Table~\ref{table:EML-res}. $A_V$ is calculated using the Balmer decrement (H$\alpha$/H$\beta$) and assuming a Milky Way extinction law. For the GTC data we take the extinction derived from the FORS spectra due to the lack of H$\alpha$ and because H$\gamma$ is affected by a skyline. Metallicities are determined using the strong line parameters N2 and O3N2 taking a recent calibration done by \citet{mar13} using $T_{\rm e}$-based abundances and, where not available, $R_{23}$ as detailed in \citet{kew07}. The SFR is calculated from H$\alpha$ using the conversion of \citet{ken98} and using extinction corrected H$\alpha$ fluxes. SFRs for the GTC data where H$\alpha$ is not present have been derived from H$\beta$ (extinction corrected) taking a ratio of 2.85 for H$\alpha$/H$\beta$ (case B recombination of hydrogen). The ionisation parameter $U$ is derived from [\ion{O}{ii}] and [\ion{O}{iii}] \citep{dia00}.

\begin{table*}[ht!]
\caption{Emission line fluxes in the extracted regions, fluxes are given in 10$^{-17}$ erg\,cm$^{-2}$\,s$^{-1}$. Values are as measured from the spectrum  and not corrected for extinction.} 
\label{table:EML}  
{\small
\begin{center}                        
\begin{tabular}{l | c | c c c c | c c c}       \hline  \hline            
{}     	&X-shooter&  GTC & GTC & GTC & GTC & FORS & FORS & FORS\\
{} &OT site & OT site & core & arm & all & OT site & core & arm\\
\hline                        
H$\alpha$ & 7.48$\pm$0.7	& ---	& ---	&---	& --- &11.4$\pm$3.4	& 27.7$\pm$1.22 & 4.73$\pm$0.83\\
H$\beta$& 1.62$\pm$0.5   &2.43$\pm$0.09	& 6.57$\pm$0.11	&3.02$\pm$0.08&13.4$\pm$0.44	&2.53$\pm$1.8	&  6.53$\pm$0.44& 1.23$\pm$0.19\\
H$\gamma$& ---   & 1.34$\pm$0.07	& 2.55$\pm$0.08	& 1.36$\pm$0.05& 6.0$\pm$0.64	&1.94$\pm$0.22	&2.74$\pm$0.42 & 5.56$\pm$0.95\\
\,[\ion{O}{ii}]$\lambda$3727& 2.66$\pm$0.39  &7.88$\pm$0.09&	18.5$\pm$0.19&8.02$\pm$0.18& 37.9$\pm$0.76	&6.29$\pm$0.37& 17.6$\pm$0.39 & 4.37$\pm$0.24\\
\,[\ion{O}{ii}]$\lambda$3730& 2.54$\pm$0.24  & (blended)	&(blended)	&(blended)	&(blended)	&(blended) &(blended)&(blended)\\
\,[\ion{O}{iii}]$\lambda$5008& 1.52$\pm$0.26  &	 2.17$\pm$0.10&4.48$\pm$0.11	&2.20$\pm$0.08& 8.90$\pm$0.47	&2.06$\pm$.17	& 5.26$\pm$0.38 & 1.10$\pm$0.33\\
\,[\ion{N}{ii}]$\lambda$6585& 1.71$\pm$0.31  &---	 &---	&--- & ---	&2.10$\pm$0.38& 7.94$\pm$1.04 & 2.84$\pm$0.69\\
\,[\ion{S}{ii}]$\lambda$6718&  1.74$\pm$0.42 &---	&---	&---& ---	&4.72$\pm$0.49& 6.11$\pm$0.60& 0.44$\pm$0.13\\
\,[\ion{S}{ii}]$\lambda$6733&  0.88$\pm$0.27 &---	&---	&---& ---	&2.46$\pm$0.44 & 4.86$\pm$0.84 & 0.83$\pm$0.14\\ \hline
\end{tabular}    
\end{center}
}
\end{table*}

\begin{table*}[ht!]\begin{footnotesize}
\caption{Results from the spatially resolved analysis. The errors displayed here include both the uncertainty from the method and the statistical one. The extinction used for the GTC data, displayed in parenthesis, is derived from the FORS spectra, as explained in the text.
} 
\label{table:EML-res}      
\begin{center}                        
\begin{tabular}{l | c | c c c c | c c c}       \hline \hline      

{}     	& X-shooter &  {} & GTC	& {}  & & &   FORS & \\
{}	& OT site & OT site & core & arm & all & OT site & core & arm \\
\hline                        
$A_V$ \,[mag]& 	1.39$\pm$0.2	& (1.33) & (1.14)		& (0.87)	& (1.2)	& 1.33$\pm$0.3	&1.14$\pm$0.2		& 0.87$\pm$0.2 \\
\small{12+log(O/H) ($R_{23}$)}& 7.94/8.98		&8.32/8.76 	& 8.20/8.86		&8.13/8.88 	& 8.22/8.85	&8.16/8.88	& 8.17/8.87	& 8.26/8.77 \\
\small{12+log(O/H) (N2)}& 	8.45$\pm$0.17	& ---	& 	---	& ---	&  --- & 	8.40$\pm$0.16	&	8.49$\pm$0.16	& 8.64$\pm$0.17 \\
\small{12+log(O/H) (O3N2)}&   8.41$\pm$0.18	& ---	& 	---	& ---	& --- &	8.40$\pm$0.18	&	8.44$\pm$0.18	& 8.50$\pm$0.18 \\
SFR H$\alpha$ \,[$M_\odot$ yr$^{-1}$]& 0.70$\pm$0.07	& (0.94)	& 	(2.22)	&(1.03) 	& (4.85) & 1.01$\pm$0.3	&	2.19$\pm$0.5	& 0.30$\pm$0.05 \\
$M_*$ [$M_\odot$] & --- & 5.2$\times$10$^8$  	& 9.8$\times$10$^8$	&2.3$\times$10$^8$ & 1.7$\times$10$^9$	& --- &1.3$\times$10$^9$	& 3.4$\times$10$^8$	\\
SSFR [Gyr$^{-1}$] &	--- & 0.96	&2.26	& 4.47 &2.85	& ---&1.68	& 0.88 \\
log $U$ ($O_{23}$)& 	--3.50	&--3.51	& --3.54		& --3.49	& --3.56& --3.46	& --3.47		&--3.49 \\
\hline                                 
\end{tabular}
\end{center}
\end{footnotesize}
\end{table*}

For the metallicity measurement we use the O3N2 calibrator whenever possible, as it is the most reliable one taking into account all the lines at our disposal. $R_{23}$ is known to have several critical problems (and should only be used when nothing else is available), mainly due to the requirement of using another diagnostic to break the $R_{23}$ degeneracy plus the large scatter at the turn-over between the two branches. The two calibrators, N2 and O3N2, give a similar metallicity and thus we discard the $R_{23}$ values for the metallicity assessment. Nevertheless, we report the values for the sake of comparison.

The metallicity of the host is similar throughout the system with a value of $\sim 0.5$ solar at the GRB site and the core, and marginally higher ($\sim$0.7 solar) at the side opposite the core of the GRB region, depending on the calibrator (these values are slightly lower than the ones obtained by \citealt{cuc13}, which is mostly due to the use of a different calibrator). This would imply a flat or positive metallicity gradient, while most spiral galaxies have  lower metallicity in the outskirts compared to the core, due to the evolution of the star formation from the core outwards. This could be due to the obvious interaction with another galaxy leading to a mixing of the gas. The metallicity is higher than for most long GRB hosts, with the exception of a few highly extinguished ones \citep[e.g.][]{kru12, per13, lev10}. The star-formation rate is moderate throughout the host with a total SFR of $\sim$ 4.8 $M_\odot$ yr$^{-1}$. This value is slightly higher than the one calculated by \citet{cuc13}, but is consistent within our different spectra and even with the one derived from the photometric fit. [\ion{O}{iii}]$\lambda$5008 is rather weak compared to [\ion{O}{ii}]$\lambda$3727 and also weaker than H$\beta$ which implies a smaller fraction of young stars providing a hard, ionising radiation field.

In the following we compare the properties of the three different extracted regions to those of other long and short GRBs as well as field galaxies from the SDSS (DR9). For the analysis of the three regions we take FORS data for the core and the part opposite the core and from X-shooter for the GRB site and corrected for extinction (if applicable). Emission line measurements as well as SFRs and masses for long and short GRB hosts were taken from the GHostS database \citep[\texttt{http://www.grbhosts.org};][]{sav09}. 

\begin{figure}[ht!]
   \centering
   \includegraphics[width=\hsize]{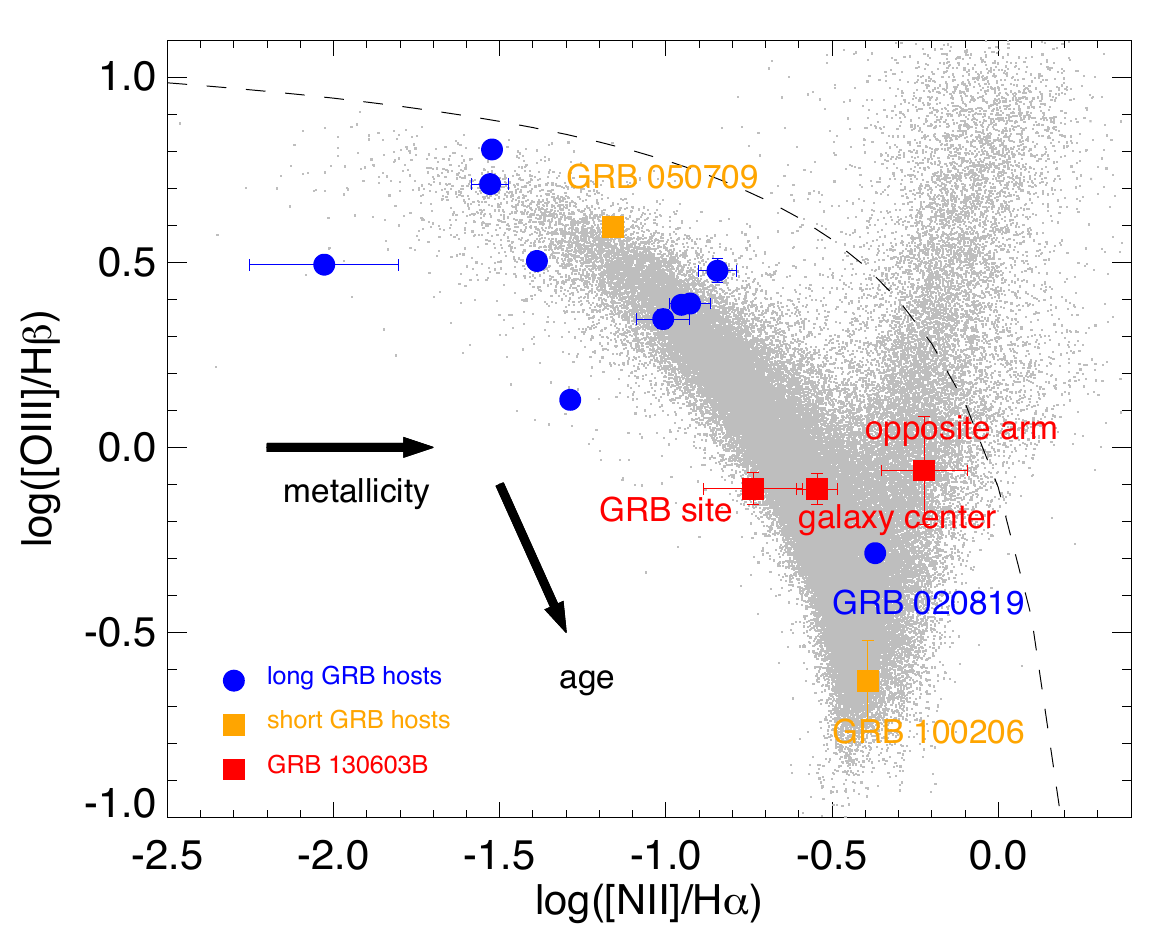}
      \caption{BPT diagram using [\ion{N}{ii}]/H$\alpha$ and [\ion{O}{iii}]/H$\beta$ in different parts of the short burst host as well as a number of other published other long (dots) and short (squares) GRB hosts (data from the GHostS database: \texttt{http://www.grbhosts.org}, see text). Grey dots are galaxies from the SDSS DR9 with a S/N or at least 10 in all emission lines used. The dashed line represents the separation between normal \ion{H}{ii} regions and AGN dominated emission. Evolution models \citep[e.g.][]{dop06} roughly indicate increasing metallicity and age as indicated by the arrows.
              }
         \label{Fig:BPT}
   \end{figure}

We plot the different regions in the ``Baldwin, Phillips and Terlevich'' \citep[BPT,][]{bal81} diagram (Fig.~\ref{Fig:BPT}) which can distinguish between regions ionised by young stars and those by AGN activity, and compare them to SDSS galaxies as well as other long and short GRB hosts. The GRB site and the core of the galaxy occupy regions consistent with average field galaxies from the SDSS and are distinctively different from long GRB hosts (excluding the very extinguished host of GRB 020819). The region occupied by long GRB hosts hint towards lower ages and metallicities from evolutionary models \citep{dop06} than the values of GRB 130603B.

\begin{figure}[ht!]
   \centering
   \includegraphics[width=\hsize]{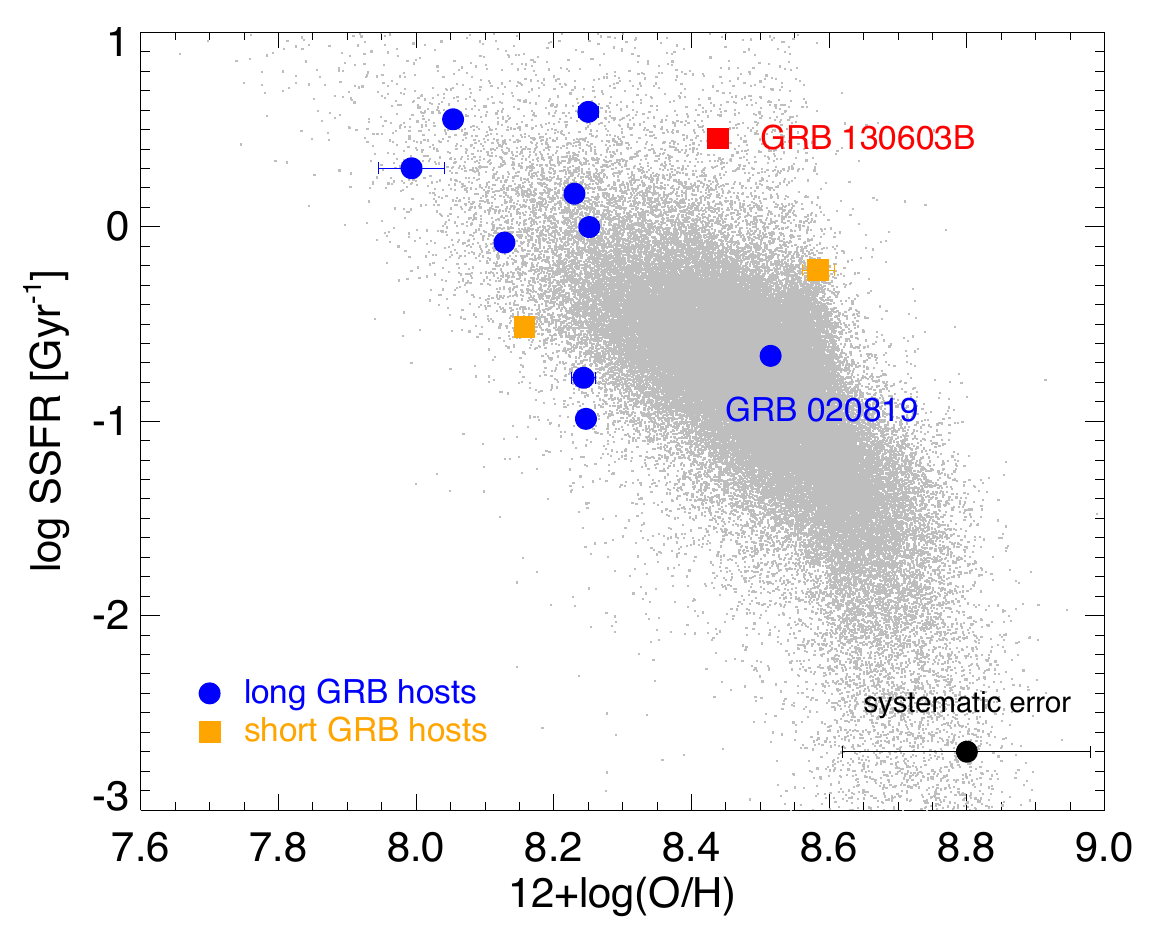}
         \caption{Metallicity versus mass-weighted star-formation rate for long and short GRB hosts (from the GHostS database, see text) compared to the distribution of SDSS galaxies. For GRB 130603B we take the average of the entire galaxy. Metallicities were obtained via the O3N2 parameter \citep{mar13}, masses from stellar population modelling. 
              }
         \label{Fig:metalssfr}
   \end{figure}
   
\begin{figure}[ht!]
   \centering
   \includegraphics[width=\hsize]{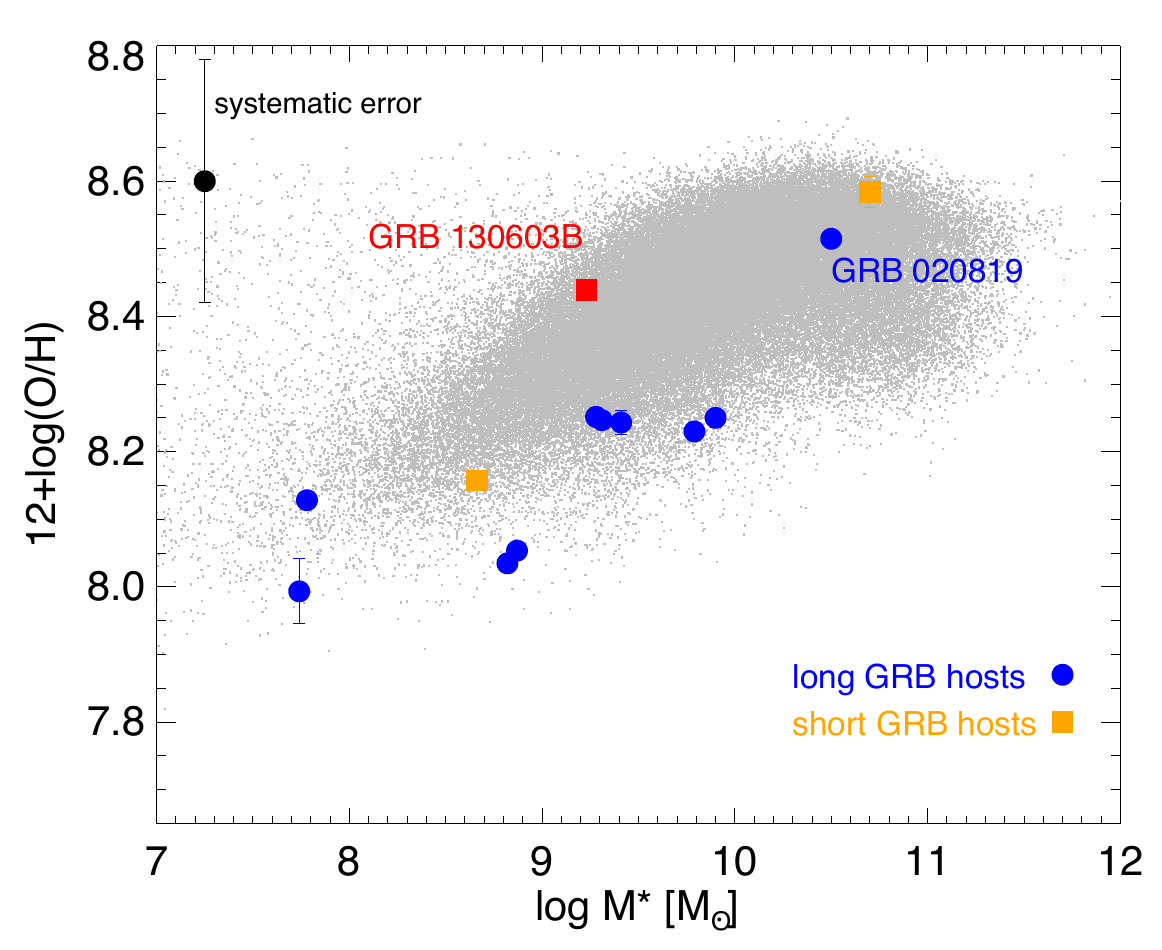}
      \caption{Mass-metallicity relation for long and short GRBs compared to SDSS DR9 galaxies and the galaxy average for GRB 130603B. Metallicities were obtained via the O3N2 parameter \citep{mar13}, masses from stellar population modelling, data for GRB hosts are from the GHostS database (see text).}
         \label{Fig:massmetal}
   \end{figure}

Furthermore, we compare the metallicity vs. specific star-formation rate (SFR weighted by the mass of the galaxy) (Fig.~\ref{Fig:metalssfr}) and the mass-metallicity relation (Fig.~\ref{Fig:massmetal}) of the same long and short bursts to SDSS galaxies. For this we use the value for the entire galaxy, not of the specific regions, and assume a metallicity from O3N2 of 12+log(O/H) = 8.44. The host of GRB 130603B has a somewhat higher metallicity than expected for a galaxy of this mass, but also a slightly higher SFR. This could both be due to a starburst at the time of the tidal interaction which is now declining but has already lead to an over-enrichment with heavier elements. Overall, the properties of the GRB site differ from the typical observations for long GRB progenitor populations, but areconsistent with the typical short GRB progenitor theory.

\section{Conclusions}

The detection of absorption lines along the line-of-sight enabled us for the first time to securely link a prototypical short GRB to its host galaxy, undoubtedly prove its cosmological origin and draw conclusions on the progenitor from its properties \citep[see also the possible kilonova detection in][]{tan13,ber13} and those of its immediate environment, which are consistent with a compact binary progenitor. The star formation within the host, location of the burst on top of the tidally disrupted arm, strong absorption features and large line of sight extinction indicate that the GRB progenitor was probably not far from its birth place, favouring merger models with a short delay time or a low natal kick velocity. GRB\,130603B has opened the door to a new era of detailed studies on short GRBs where larger samples will reveal more information on the diversity of their progenitors and establish the differences and analogies with respect to other cosmic explosions.

\begin{acknowledgements}

Based on observations collected at CAHA/Calar Alto, GTC/La Palma, WHT/La Palma, NOT/La Palma, TNG/La Palma, VLT/Paranal, UKIRT/Mauna Kea, Gemini-North/Mauna Kea.
The research activity of AdUP, CT and JG is supported by Spanish research project AYA2012-39362-C02-02.
AdUP acknowledges support by the European Commission under the Marie Curie Career Integration Grant programme (FP7-PEOPLE-2012-CIG 322307).
AR, AJvdH and RAMJW acknowledge support from the European Research Council via Advanced Investigator Grant no. 247295.
JG is supported by the Unidad Asociada IAA-CSIC\_ETSI-UPV/EHU and the Ikerbasque Foundation for Science.
RGB acknowledges support from MICINN AYA2010-15081.
TK acknowledges support by the European Commission under the Marie Curie Intra-European Fellowship Programme
BMJ and JPUF acknowledge support from the ERC-StG grant EGGS-278202.
The Dark Cosmology Centre is funded by the DNRF.

\end{acknowledgements}


\bibliographystyle{aa}
\bibliography{130603B_AA}

  \Online

\begin{appendix}

\section{Observations}

\subsection{Photometric observations}

We conducted comprehensive photometric observations of the afterglow and host galaxy of GRB\,130603B from a range of facilities described below. These data include the original discoveries of the afterglow, track the afterglow behaviour over the first $\sim 24$ hours and place stringent limits at later times. We also include the ultraviolet and optical observation from the UVOT telescope onboard the \textit{Swift} satellite \citep{dep13}. 
Details of each of the observations are given below. 

In order to separate the contribution of the host galaxy from the afterglow light, we performed image subtraction as described in the next section, resulting in the photometry provided in Table~\ref{table:photlog}. The analysis performed on the \textit{Swift}/UVOT data is significantly different from the ground based photometry. Due to this, we display the results in separate tables (Tables~\ref{table:uvot} and \ref{table:uvot2}).

\begin{figure}[ht!]
   \centering
   \includegraphics[width=\hsize]{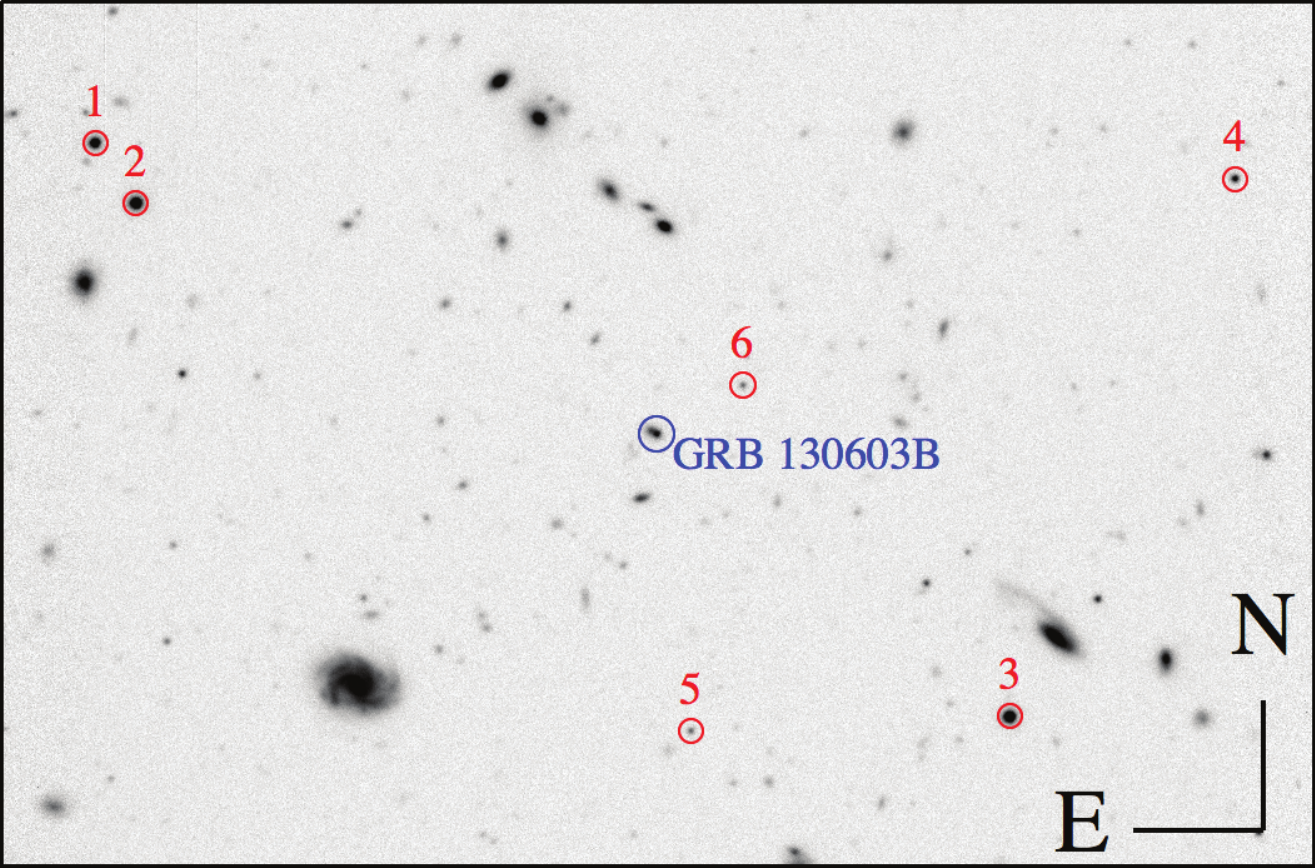}
      \caption{Finding chart showing the location of the afterglow (in blue) and of the stars used for the photometry (in red), as indicated in Table~\ref{table:std}. The field of view is $3.8^{\prime}\times2.5^{\prime}.$
              }
         \label{Fig:fc}
   \end{figure}

\begin{table}[htdp]
{\small \begin{center}
\begin{tabular}{ccccc}
\hline
\hline
Star & $g$ & $r$ & $i$ & $z$ \\
\hline
1 & 20.89 $\pm 0.02$ & 19.44 $\pm$ 0.02 & 18.35 $\pm$ 0.02 & 17.79 $\pm$ 0.02 \\
2 & 20.24 $\pm 0.02$ & 18.83 $\pm$ 0.02 & 17.90 $\pm$ 0.02 & 17.41 $\pm$ 0.02 \\
3 & 20.35 $\pm 0.02$ & 18.87 $\pm$ 0.02 & -    & -    \\
4 & 22.33 $\pm 0.04$ & 21.01 $\pm$ 0.02 & 19.81 $\pm$ 0.03 & 19.22 $\pm$ 0.03 \\
5 & 23.08 $\pm 0.04$ & 22.27 $\pm$ 0.04 & 21.96 $\pm$ 0.03 & 21.81 $\pm$ 0.03 \\
6 & 23.48 $\pm 0.05$ & 22.33 $\pm$ 0.04 & 21.89 $\pm$ 0.03 & 21.62 $\pm$ 0.03 \\
\hline
\end{tabular}
\end{center}}
\caption{Photometric comparison stars used for our optical ($griz$) photometry of GRB~130603B.}
\label{table:std}
\end{table}%

\subsubsection{{\it Swift}-UVOT observation and Analysis} 

{\it Swift's} Ultraviolet Optical Telescope \citep[UVOT;][]{rom00,rom04, rom05} began settled observations of the burst approximately 62 s after the BAT trigger. A faint source at the location of the afterglow was detected in all 7 UVOT filters.

The photometry was obtained from the image lists using the {\it Swift} tool {\tt uvotsource}. When the source is faint and the count rate low, it is more accurate to use a small source aperture to extract the photometry \citep{poo08,bre10}. Therefore we used a 3$^{\prime\prime}$ source region to extract the count rate of the source. In order to be consistent with the UVOT calibration, these count rates were then corrected to 5$^{\prime\prime}$ using the curve of growth contained in the calibration files. Background counts were extracted using a circular region of radius 20$^{\prime\prime}$ from a blank area of sky situated near to the source position. The background subtracted count rate was then converted to AB magnitudes using the UVOT photometric zero points \citep{bre11}. The analysis pipeline used software HEADAS 6.13 and UVOT calibration 20130118. To improve the signal to noise, the count rates in each filter were binned using $\Delta t/t = 0.5$. The result of this photometry is given in Table~\ref{table:uvot}.

\begin{table}[htdp]
\begin{footnotesize}
\begin{center}
\begin{tabular}{lllll}
\hline\hline
Mid $t-t_0$ & Half bin width & AB Mag	& Mag Error		& Filter	\\
(s)	&	(s)	 &		&	      		&		\\
\hline		
137	&	75	&	21.56	&	+0.35,	-0.26	&	{\it white}	\\	
651	&	97	&	$>$20.47	&	--      	&	{\it white}	\\
5599	&	818	&	21.60	&	+0.21,	-0.18	&	{\it white}	\\
51526	&	170	&	21.99	&	+0.34,	-0.26	&	{\it white}	\\
696	&	91	&	$>$18.15	&	--            	&	{\it V}	\\							
7745	&	2552	&	$>$19.57	&	--            	&	{\it V}	\\
37325	&	9331	&	$>$20.18	&	--            	&	{\it V}	\\
627	&	96	&	$>$19.84	&	--	        &	{\it B}	\\
8450	&	3874	&	20.81	&	+0.26,	-0.21	&	{\it B}	\\
42451	&	8900	&	$>$20.82	&	--       	&	{\it B}	\\
487	&	212	&	$>$20.68	&	--       	&	{\it U}	\\
5189	&	818	&	$>$20.80 	&	--       	&	{\it U}	\\
80635	&	40241	&	$>$21.63  &	--       	&	{\it U}	\\
134782	&	9495	&	$>$22.31	&	--       	&	{\it U}	\\
664	&	10	&	$>$19.60 	&	--       	&	{\it UVW1}	\\
4984	&	818	&	$>$21.42	&	--       	&	{\it UVW1}	\\
29117	&	11270	&	$>$22.35	&	--       	&	{\it UVW1}	\\
112413	&	32582	&	23.05	&	+0.28,	-0.22	&	{\it UVW1}	\\
640	&	10	&	$>$20.54	&	--       	&	{\it UVM2}	\\
4778	&	819	&	$>$21.40	&	--      	&	{\it UVM2}	\\
27494	&	11986	&	22.54	&	+0.32,	-0.25	&	{\it UVM2}	\\
677	&	97	&	$>$20.68	&	--              &	{\it UVW2}	\\
5772	&	783	&	$>$21.8 	&	--       	&	{\it UVW2}	\\
36426	&	9338	&	$>$22.66	&	--       	&	{\it UVW2}	\\						
270445	&	17351	&	$>$23.12	&	--       	&	{\it UVW2}	\\			
\hline
\end{tabular}
\end{center}			
\end{footnotesize}
\caption{UVOT observations of the GRB 130603B afterglow. Column 1: mid-time of exposure, in seconds, since BAT trigger; column 2: half bin width; column 3: AB magnitude; column 4: magnitude error; column 5: UVOT filter. When the signal to noise is below 3, a 3-$\sigma$ upper limit is given.}
\label{table:uvot}
\end{table}%

To improve the signal to noise of the UVOT data, a single filter light curve was computed from the 7 UVOT filters (see Table~\ref{table:uvot2}). The count rates in each filter were normalised to the \textit{U} filter and then the resulting light curve was binned using $\Delta t/t$ = 1.0 \citep{oat09}.

\begin{table}[htdp]
\begin{footnotesize}
\begin{center}
\begin{tabular}{lllllllllll}
\hline\hline
Mid $t-t_0$ 	& Half bin width & AB Mag   & Mag Error	& Filter	\\
(s)		&	(s)	&		&      			&			&		\\
\hline																														
137		& 75		&	21.54	&	+0.35,	-0.26	&  {\it U}	\\									
531		& 256	&	21.59	&	+0.40,	-0.29	&  {\it U}	\\			
7128		& 3168	&	21.62	&	+0.16, 	-0.14	&  {\it U}	\\
22966	& 11331	&	21.50	&	+0.16, 	-0.14&  {\it U}	\\
76639	& 38058	&	22.26	&	+0.20, 	-0.17&  {\it U}	\\
203606	& 84191	&	22.92	&	+0.32, 	-0.25&  {\it U}	\\								
\hline
\end{tabular}
\end{center}
\caption{UVOT observations of the GRB 130603B afterglow, combining the different filters and normalised to the {\it U} filter in to improve the signal-to-noise ratio.}
\label{table:uvot2}
\end{footnotesize}
\end{table}%

\subsubsection{1.23m CAHA}

The  1.23m  telescope  is  an   f/8  optical  system  located  in  the observatory of Calar Alto,  Almer\'{\i}a, Spain. The observations were carried out in the Johnson \textit{V}-band and a DLR-MKIII camera.  The DLR-MKIII camera is based  on a 4k$\times$4k e2v CCD231-84-NIMO-BI-DD detector. The  field of  view achieved is  21.5$^{\prime}$ $\times$  21.5$^{\prime}$. The data were acquired using a 2$\times$2 binning read-out, providing a pixel size of $0.63^{\prime\prime}$ pix$^{-1}$.

\subsubsection{NOT}

Observations in \textit{r}-band were obtained with the MOSCA instrument at the Nordic Optical Telescope (NOT) beginning 0.25 days after the burst discovery. Additional observations were obtained on 5 June 2013 and used as a template for subtraction. 

\subsubsection{WHT}

We observed the location of GRB\,130603B with the William Herschel Telescope (WHT) on La Palma on two occasions on 3 June 2013 and 6 June 2013, with the first observations beginning $\sim$ 0.25 days post burst. At each epoch we obtained observations in $g$, $i$ and $z$ using the Auxiliary Port Camera (ACAM). The images were debiased and flat-fielded in the standard way within IRAF, with the second epoch subsequently subtracted from the first to provide clean subtractions for photometric observations. 

\subsubsection{GTC}

Optical imaging was carried out with the Gran Telescopio Canarias (GTC), a 10.4 m telescope located at  the observatory of Roque de los Muchachos  on La Palma (Canary Islands, Spain), and equipped with the OSIRIS instrument. The observations were obtained with 2$\times$2 binning mode, yielding a pixel scale of 0.26$^{\prime\prime}$ pix$^{-1}$ and a field of view $7.8^\prime \times 8.5^\prime$.

\subsubsection{VLT}

We observed the field of GRB\,130603B with the Very Large Telescope (VLT) utilising both optical (FORS) and infrared (HAWK-I) observations. Our first photometric observations were obtained as part of the spectroscopic acquisition. 
Each observation was reduced with the appropriate instrument pipeline via {\tt esorex}, to create dark/bias subtracted, flat-fielded images which were subsequently combined into final images. Photometric calibration in the optical is given relative to SDSS for \textit{g}-band observations and to our own \textit{V}-band calibration for the \textit{V}-band acquisition. 

\subsubsection{Gemini}

We obtained two epochs of optical ($griz$) imaging of GRB 130603B with Gemini-N using the GMOS-N instrument over the first $\sim$ 2 days post burst. An additional epoch was obtained with Gemini-S (using GMOS-S which has an almost identical instrument setup) on 15 June 2013. Image subtraction was performed between the second epoch of Gemini-N and the Gemini-S observations, confirming no transient emission over this timescale. Given this, and the relative ease (and cleanliness) of the subtractions we use the second epoch of Gemini-N observations as a template for subtraction for the first. 

\subsubsection{TNG}

We imaged the field of the short GRB 130603B with the Italian 3.6m Telescopio Nazionale Galileo (TNG), located in La Palma, Canary Islands. Optical observations with the $r$ and $i$ SDSS filters were carried out with the DOLoReS camera on 2013 Jun 16 and 24. All nights were clear, with seeing in the range 1.1$^{\prime\prime}$-1.2$^{\prime\prime}$. However, on June 24 the observations were performed under bright Moon conditions. 

Image reduction was carried out following the standard procedures: subtraction of an averaged bias frame, division by a normalised flat frame. Astrometry was performed using the USNO B1.0 catalogue\footnote{\texttt{http://www.nofs.navy.mil/data/fchpix}}. Aperture photometry was made with the DAOPHOT task for all objects in the field. The photometric calibration was done against the SDSS catalogue. The flux limits on the presence of an optical transient were estimated using small apertures centred on the optical afterglow position. In order to minimise any systematic effect, we performed differential photometry with respect to a selection of local, isolated an not-saturated reference stars visible in the field of view.

\subsection{Spectroscopy}

Spectroscopic observations were obtained using different telescopes and instruments as shown in the observing log (Table~\ref{table:speclog}) and detailed in the following paragraphs. The slits of the spectrographs were positioned at different position angles, providing different coverage of the underlying host galaxy. A diagram with the slit position of the most relevant observations is shown in Fig.~\ref{fig:finder}.

\subsubsection{GTC/OSIRIS}

We  acquired  spectroscopic data using the OSIRIS imager and spectrograph at the GTC. The slit was oriented with a position angle (measured from North to East) of 51$\degree$. In order to determine the host contribution of the first observation and to be able to isolate the afterglow emission we performed a second observation on day 5, when the afterglow contribution was negligible. The slit was positioned at the same angle and the seeing conditions were similar. 

\subsubsection{VLT/X-shooter}

We acquired a medium resolution spectrum with the X-shooter spectrograph mounted at the ESO/VLT, covering the range from 3000 to 24800\,\AA. Observations began at
00:00:36 on 04 June, 2013 and comprise of 4 nodded exposures, having an individual
integration time of 600 s in the UVB, VIS and NIR arms (the mid exposure time
is 8.555 hr post burst; \citealt{xu13}). Due to technical problems with the atmospheric
dispersion correctors, we set the position angle to the parallactic angle (being
170.1$\degree$ at that epoch), see Fig. \ref{fig:finder}, i.e. the X-shooter spectrum does
not cover other parts of the host galaxy. The slit widths were 1.0$^{\prime\prime}$, 0.9$^{\prime\prime}$, and
0.9$^{\prime\prime}$ in the UVB, VIS, and NIR, respectively. For this given instrument setup, the resolution is 5100, 8800, and 5300 in the UVB, VIS and NIR, respectively.

VLT/X-shooter data were reduced with the X-shooter pipeline v2.0 \citep{gol06}\footnote{\texttt{http://www.eso.org/sci/software/pipelines/}}.
The wavelength binning was chosen to be 0.2 \AA pix$^-1$ in the UVB and VIS, and a 0.5
\AA pix$^-1$ binning in the NIR. All spectra were flux calibrated with the spectrophotometric
standard star LTT3218 and scaled with our photometric data to correct for slit losses.
Furthermore, data were corrected for Galactic extinction ($E(B-V)=0.02$ mag). We
transformed the wavelength solution to vacuum. 
No attempt was made to correct for telluric absorption lines. This has no implications on our analysis.

\subsubsection{VLT/FORS2}

We obtained further spectroscopy with the Focal Reducer and Spectrograph (FORS2) on the VLT.
Three spectroscopic exposures of each 600 seconds exposure time were obtained, starting at 
23:57:43 UT on 3 June 2013 (i.e. midtime of the observation was 8.40 hr after burst),
at average airmass of 1.4.  We used a 1$^{\prime\prime}$ slit width and the 300V grism without order 
separation filter (to extend the wavelength range, at the expense of some order overlap). The CCDs 
were binned $2\times2$, resulting in a spatial scale along the slit of 0.25$^{\prime\prime}$ pix$^{-1}$. 
Data were reduced using IRAF routines, using calibration data (flatfields and arc lamp exposures) 
taken the same night, the three exposures were combined before source extraction. We measure 
a FWHM spectral resolution $R=590$ at 7000 \AA. The slit position angle (58$\degree$) was chosen such that 
the slit covers both the afterglow and the host nucleus, giving the largest possible spatial separation of
the afterglow spectral trace from that of the brightest part of the host galaxy. We are therefore able to 
extract the afterglow spectrum and host spectrum with minimised contamination from one to the other,
while using both traces to obtain an accurate trace position fit. We used an observation of the 
spectrophotometric standard star BD 25 4655 to correct for instrument response, and photometric 
observations were used to bring the data to an absolute flux scale.

\subsubsection{WHT/ACAM}

We obtained spectroscopy using ACAM on the WHT immediately following the identification of the 
afterglow in the imaging data. A single exposure with 900 s of integration was obtained, 
centred on the afterglow, at an average airmass of 1.9, starting at 23:54:36 UT on 3 June 
2013 (i.e. the mid time of the observation was 8.22 hr after the burst). The slit position angle was set to parallactic.  
Data were reduced using IRAF routines, using calibration data (flatfields and arc lamp exposures) 
taken the same night. ACAM uses a 400 lines/mm volume phase holographic grating, and we
used a 1$^{\prime\prime}$ slit width; we measure a FWHM spectral resolution $R=530$ at 7000 \AA. The 
trace shows a strong continuum. Several emission lines from the host are also detected, somewhat offset from the 
afterglow trace photocenter and at a redshift consistent with the data from VLT and GTC; absorption lines are 
only marginally significant.

\section{The afterglow}

\subsection{Photometry of the afterglow}

To obtain photometric measurements of the afterglow in each band we performed, when possible, image subtraction with the public ISIS code \citep{ala00}. For clean subtractions we selected a later time (afterglow free) image from each telescope as a template and subtract this from the earlier data. Photometric calibration of these subtracted images was obtained by creating artificial stars of known magnitude in the first image, with the errors estimated from the scatter in a large number of apertures (of radius approximately equal to the seeing) placed within the subtracted image. The placement of artificial stars close to the limiting magnitude within the image confirms that these can be recovered, and so the given limiting magnitudes are appropriate. However, we do note that the limiting magnitudes are based on the scatter in photometric apertures placed on the sky, not on the relatively bright regions of the host directly underlying the GRB. 
The log of optical/NIR photometry of the afterglow is given in Table~\ref{table:photlog}.

We used the stacked UVOT data (\ref{table:uvot2}), where the last epoch was subtracted from the rest, assuming that this last one had only emission from the host galaxy. These data are important, as they cover the early phase of the optical light curve and allow us to identify a flattening in the optical light curve which significantly differs from the X-ray emission.

\begin{figure}[ht!]
   \centering
   \includegraphics[width=\hsize]{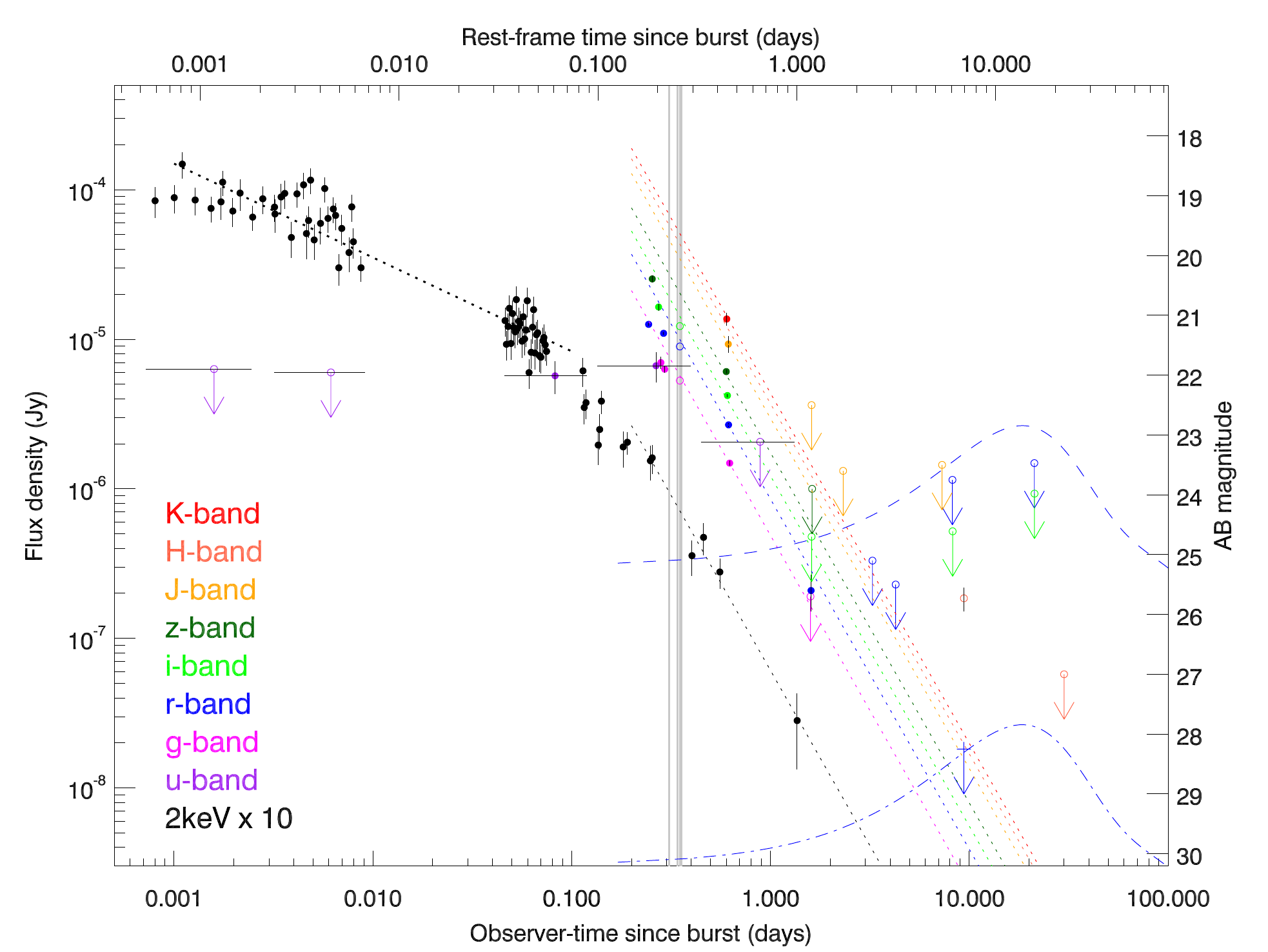}
      \caption{Light curves of GRB\,130603B, indicated detections with dots and upper limits (3-$\sigma$) with arrows. $V$-band photometry has been scaled and plotted together with the $g$-band. The vertical lines indicate the times when spectra were obtained. Dotted lines indicate the light curve fits to a power law temporal decay from 0.3 to 3 days after the burst. We include data from the literature \citep{cuc13, tan13}. The dashed blue line is the expected \textit{r}-band light curve of a supernova like SN~1998bw, the most common template for long GRBs after including an extinction of $A_V=0.9$ magnitude. The most constraining limits indicate that any supernova contribution would be at least 100 times dimmer than SN~1998bw in the $r$-band, once corrected of extinction (blue dashed-dotted line).
              }
         \label{Fig:lc1}
   \end{figure}

The X-Ray Telescope (XRT) onboard {\it Swift} began observing the GRB field on 2013 Jun 03 at 15:49:13.945 UT. We used photon counting (PC) mode data for the X-ray spectral analysis. X-ray data were obtained with \textit{Swift}/XRT using the reductions provided by the {\it Swift} Burst Analyser \citep{eva10} and transformed to 2 keV. The light curves obtained with all these observations are plotted together in Fig.~\ref{Fig:lc1}.

\begin{figure}[ht!]
   \centering
   \includegraphics[width=\hsize]{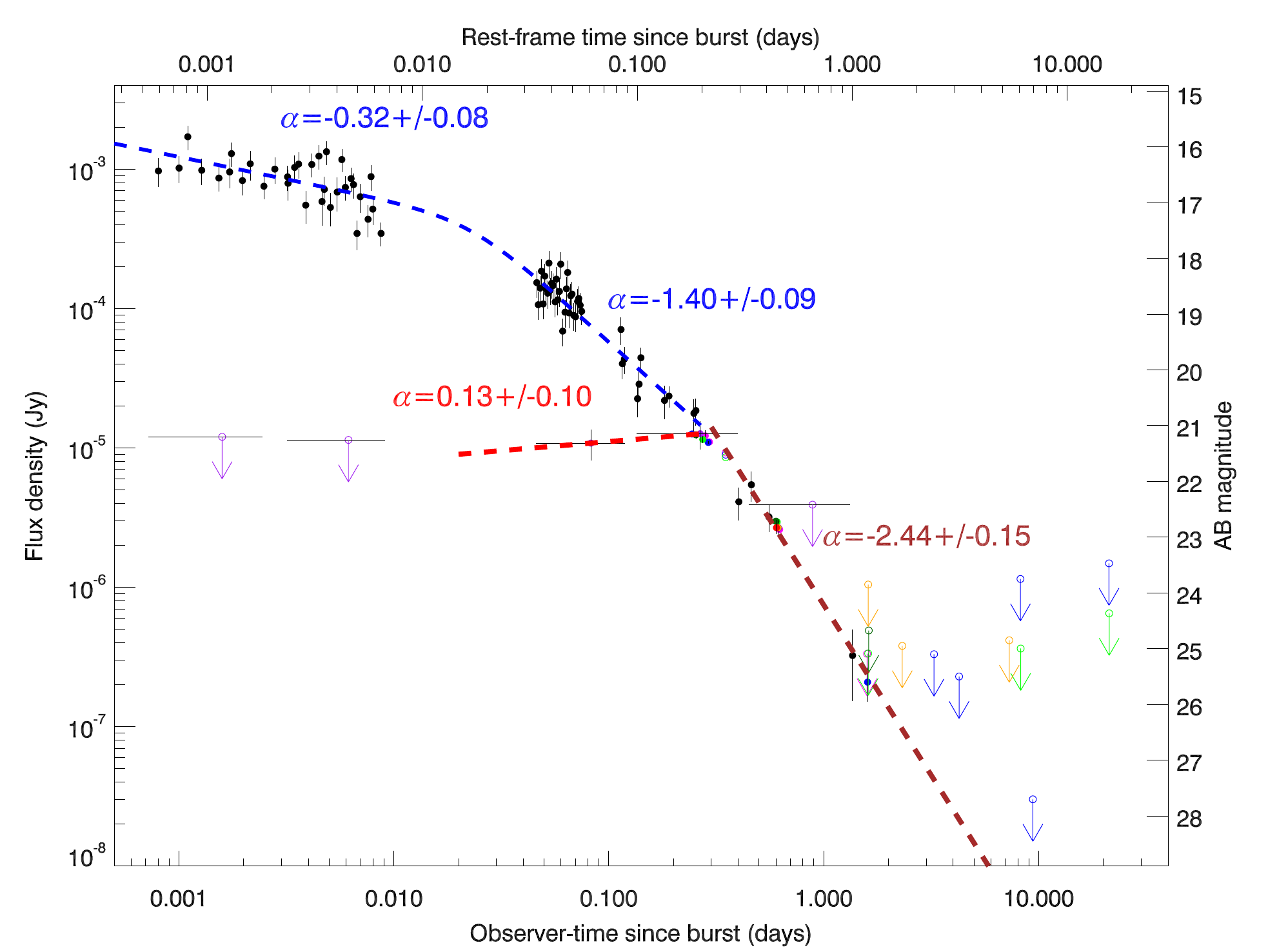}
      \caption{Light curve of GRB\,130603B, where all the bands have been scaled to the \textit{r}-band 0.3 days after the burst. The evolution of all the bands is consistent after $\sim$0.25 days, but before that the X-rays are much stronger than the optical which seem to reach a maximum at around 0.2 days. Overplotted are the spectral slopes of the fits to some of specific segments of the light curve, where $F_{\nu}\propto t^{\alpha}$.
              }
         \label{Fig:lc2}
   \end{figure}

The X-ray light curve is characterised by a slow decay during the first 0.1 days, followed by a gradual steepening. The late optical data match reasonably well the steep late decay of the X-ray light curve, which makes the afterglow undetectable after one day even for large telescopes. There is a clear break in the optical light curve at $\sim0.25$ days before which the evolution strongly differs from the X-ray one, with the optical being much flatter than the X-rays or even consistent with a brightening until 0.2 days. A direct comparison of optical and X-ray light curves is shown in Fig.~\ref{Fig:lc2}.

\subsection{Spectral energy distribution of the afterglow and extinction}

In this section we aim to fit the X-ray to optical/NIR SED using the method followed in \citet{zaf11,zaf12} to derive the extinction in the line of sight of the GRB and determine some spectral parameters. The procedure is briefly explained below. 

The flux calibrated spectrum has been analysed after removing wavelength intervals affected by telluric lines and strong absorption lines. We then rebinned the spectrum in bins of approximately 8\,\AA\ by a sigma-clipping algorithm. To check the flux calibration of the X-shooter spectrum, we compare the continuum with the flux densities obtained from the extrapolation of the photometry at the time of the spectrum (mid time around 8.56 hr). 

We include the X-ray spectrum from XRT on board \emph{Swift}. We used \texttt{XSELECT} (v2.4) to extract spectral files from the event data in the 0.3--10 keV energy band. The X-ray spectrum was extracted in the time interval 10000 to 52000 s, resulting in 9.7 ks effective exposure time. The X-ray spectral file was grouped to 15 counts per energy channel. The spectra were fitted within \texttt{XSPEC} \citep[v12.8.0;][]{arn96} with a Galactic-absorbed power-law model, with absorption from the Galactic neutral hydrogen column density (fixed to $1.93 \times 10^{20}$ cm$^{-2}$; \citealt{kal05}), and absorption from the GRB host galaxy. The best fit parameters are the photon index $\Gamma=2.26^{+0.25}_{-0.18}$ and the intrinsic column density $N_{\rm H_X}=1.34^{+0.14}_{-0.16} \times 10^{21}$ cm$^{-2}$. The resulting fit is good with $\chi^2_{\rm red}=0.99$ for 11 degrees of freedom. 

We try to fit the data using different extinction curves, namely SMC, LMC, and MW, together with single and broken power-law models. The best fit to the X-ray to optical/NIR SED results in an extinction of $A_V=0.86\pm0.15$ mag and an SMC extinction law. The SED is well fitted with a broken power-law with the optical slope $\beta_{\rm O}=-0.65\pm0.09$, X-ray slope $\beta_{\rm X}=-1.15\pm0.11$, and a spectral break at log $\nu_{\rm break/Hz} = 15.98\pm0.76$. The resulting fit is good with $\chi^2_{\rm red}=1.05$ for 2615 degrees of freedom. We also fit the data with a single power-law model resulting in a poor fit with best fit parameters of $A_V=0.65\pm0.13$ mag and $\beta=-0.89\pm0.13$ and a $\chi^2_{\rm red}=1.68$ for 2617 degrees of freedom.

One of the key diagnostics of where GRBs explode is the absorption in their afterglow spectra, with LGRBs showing very high soft X-ray absorbing column densities ($N_{\rm H_X}$) on average, associated with the star-forming regions in which they explode \citep[e.g.][]{gal01}. Oddly, analyses of the small sample of SGRBs obtained over the past decade with {\it Swift} have shown that SGRBs can have high X-ray absorbing column densities too \citep{kop12}, and a distribution in this absorption comparable to long GRBs. This result is surprising given the very different expected environments and points to a high density environment for some SGRBs. However, these results are based on small number statistics, are biased towards GRBs closer to their host galaxies (to have firmer associations), and are contaminated with short GRBs with extended emission (50\% of the mentioned sample), so the issue is not settled.

\section{GRB\,130603B in the context of short GRBs}

We select all {\it Swift} GRBs with $T_{90}\le$2 s, using the standard classification of short GRBs \citep{kou93}, and a firm host galaxy association with a measured redshift. These selection criteria exclude short GRBs with extended emission, which may share a common progenitor to short GRBs but this has not been proven to date. The sample is given in Table \ref{SGRB_sample}. GRB\,130603B is an unambiguously short GRB with one of the shortest $T_{90}$ durations in the sample of short GRBs (with well constrained host galaxy redshifts), spectrally harder than the average short GRB in this sample and with slightly higher than average fluence. To compare with this sample we construct pseudo-bolometric light curves using the gamma-ray and X-ray data from the {\it Swift} satellite as explained below. The unabsorbed observed 0.3--10 keV light curves were downloaded from the {\it Swift} Burst Analyser using 5$\sigma$ binning \citep{eva10}. When 5$\sigma$ binning resulted in a non-detection in the BAT observations, we manually bin the BAT data using 3$\sigma$ significance bins and extrapolate each bin to 0.3--10 keV assuming a simple power-law spectral model. These data points were combined with the XRT light curves obtained from the {\it Swift} Burst Analyser. Each light curve was converted into rest frame 1--10000 keV luminosity light curves using a k-correction \citep{blo01} giving an approximation to a bolometric light curve. This procedure is also repeated for all long GRBs with photometric and spectroscopic redshifts.

\begin{figure}
\begin{center}
\includegraphics[bb= 30 40 530 700, clip,angle=270, width=\hsize]{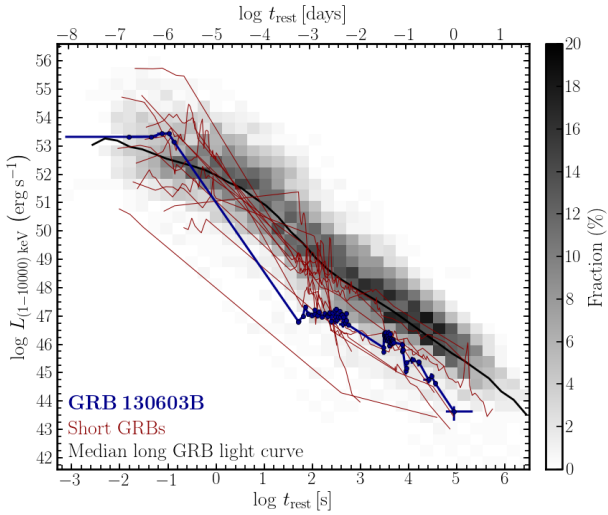}
\caption{Pseudo-bolometric light curve evolution of \textit{Swift} GRBs from the prompt to the
	afterglow phase. The density plot was built from over 280 long GRBs with redshift information.
	Its median light curve is shown in black. Overplotted are in red the 13 short GRBs from Table
	\ref{SGRB_sample} and  in dark blue GRB\,130603B. X-ray data sets naturally have orbit gaps. To correct 
	for this, we estimate the luminosity of long GRBs at missing grid points by interpolating between
	adjacent data points. 
}
\label{fig:SGRBs_v_LGRBs}
\end{center}
\end{figure}

In Fig.~\ref{fig:SGRBs_v_LGRBs} we plot the rest frame light curve of GRB\,130603B in comparison to other short GRBs and the distribution of long GRBs. The luminosities of long and short GRBs during the prompt emission should be compared with great caution. We assumed that the spectrum of all bursts are adequately described with a power law from 0.3 to 10000 keV, however most GRBs have a peak energy of a few hundred keV \citep{nav11}. Depending on the intrinsic spectrum, this can result in an uncertainty in the bolometric luminosity of about three orders of magnitude. Despite these uncertainties, it appears that short GRBs can be as luminous as the most luminous long GRBs during the prompt emission. The comparison of the X-ray afterglow is not affected by this limitation, because by the time the X-ray observations begin, the peak of the intrinsic spectrum is not at $\gamma$-ray energies anymore but in the mm/sub-mm range. The short GRB afterglows are offset from the long GRBs, as they are fainter and typically fade more rapidly but otherwise show similar behaviour \citep[consistent with the findings of e.g.][]{nys09}. GRB 130603B is consistent with the rest of the short GRB sample, as it has an average luminosity for a short GRB but is about 1.5 dex less luminous than an average long GRB and only consistent with less than 1--2\% of the long GRB sample.

\begin{table*}
\begin{center}
\begin{tabular}{cccccc}
\hline\hline
GRB     & $z$ & $T_{90}$ & $\Gamma_{\gamma}$ & Flux & References\\
        &   & (s)      &          & (10$^{-7}$ erg cm$^{-2}$ s$^{-1}$) &\\
\hline
050509B  & 0.23    & 0.024$\pm$0.009 & 1.50$\pm$0.40   & 0.2$\pm$0.1& \citealt{bar05,geh05} \\
051221A  & 0.55    & 1.4$\pm$0.2     & 1.39$\pm$0.06   & 11.6$\pm$0.4 & \citealt{cum05,sod06}\\
060801   & 1.13    & 0.5$\pm$0.1     & 0.47$\pm$0.24   & 0.8$\pm$0.1  & \citealt{sat06,cuc06} \\
061201   & 0.111   & 0.8$\pm$0.1     & 0.81$\pm$0.15   & 3.3$\pm$0.3 &  \citealt{mar06,str07} \\
070724A & 0.46    & 0.4$\pm$0.04    & 1.81$\pm$0.33   & 0.30$\pm$0.07 & \citealt{par07,ber09} \\
070809   & 0.219   & 1.3$\pm$0.1     & 1.69$\pm$0.22   & 1.0$\pm$0.1  & \citealt{kri07,per08} \\
080905A  & 0.122   & 1.0$\pm$0.1     & 0.85$\pm$0.24   & 1.4$\pm$0.2  & \citealt{cum08,row10} \\
090426   & 2.6     & 1.2$\pm$0.3     & 1.93$\pm$0.22   & 1.8$\pm$0.3  & \citealt{sat09,ant09} \\
090510   & 0.9     & 0.3$\pm$0.1     & 0.98$\pm$0.20   & 3.4$\pm$0.4  & \citealt{ukw09,mcb10} \\
100117A & 0.92    & 0.30$\pm$0.05   & 0.88$\pm$0.22   & 0.9$\pm$0.1 & \citealt{mar10,fon11}  \\
101219A & 0.718   & 0.6$\pm$0.2     & 0.63$\pm$0.09   & 4.6$\pm$0.3   & \citealt{kri10, cho11}\\
111117A & 1.31    & 0.47$\pm$0.09   & 0.65$\pm$0.22   & 1.4$\pm$0.2 & \citealt{sak11b,sak13}  \\
120804A & 1.30    & 0.81$\pm$0.08   & 1.34$\pm$0.08   & 8.8$\pm$0.5 & \citealt{bau12, ber13b}  \\
130603B & 0.356   & 0.18$\pm$0.02   & 0.82$\pm$0.02   & 6.3$\pm$0.3 & \citealt{bar13}  \\
\hline
\end{tabular}
\end{center}
\caption{The SGRB sample with redshift, typically from firm host galaxy associations, with their $T_{90}$ duration, photon index ($\Gamma_{\gamma}$) and fluence (15--150 keV).} 
\label{SGRB_sample}
\end{table*}

\section{Comparison of the Fireball model to the Magnetar model}

In the fireball model of GRBs, the late time X-ray and optical emission originates from a forward shock interacting with the surrounding medium with a reverse shock propagating back through the jet and, with well sampled optical and X-ray data, it is possible to derive the electron energy distribution and structure of the surrounding medium \citep{ree92, mes97,wij97}. 
At later times, the light curve is predicted to have an achromatic jet break which can be used to derive values such as the jet opening angle \citep[e.g.][]{kum00}. This model can be fit to GRB 130603B, assuming that the optical data are consistent with a fireball peaking at $\sim$0.15 days with a jet break at $\sim$0.25 days and a photon index of $p=2.3$. However, the X-ray emission would only be consistent with this if there is an additional component prior to $\sim$0.1 days. For the observations to be consistent with the fireball model, the peak of the emission should be reached at unusually late timescales which implies that the initial Lorentz factor was very low. Additionally, the pre-jet break optical and X-ray temporal slopes are inconsistent with the fireball model. Therefore, GRB 130603B has evidence of early-time energy injection in the X-rays. 

\begin{figure}
\begin{center}
\includegraphics[width=\hsize]{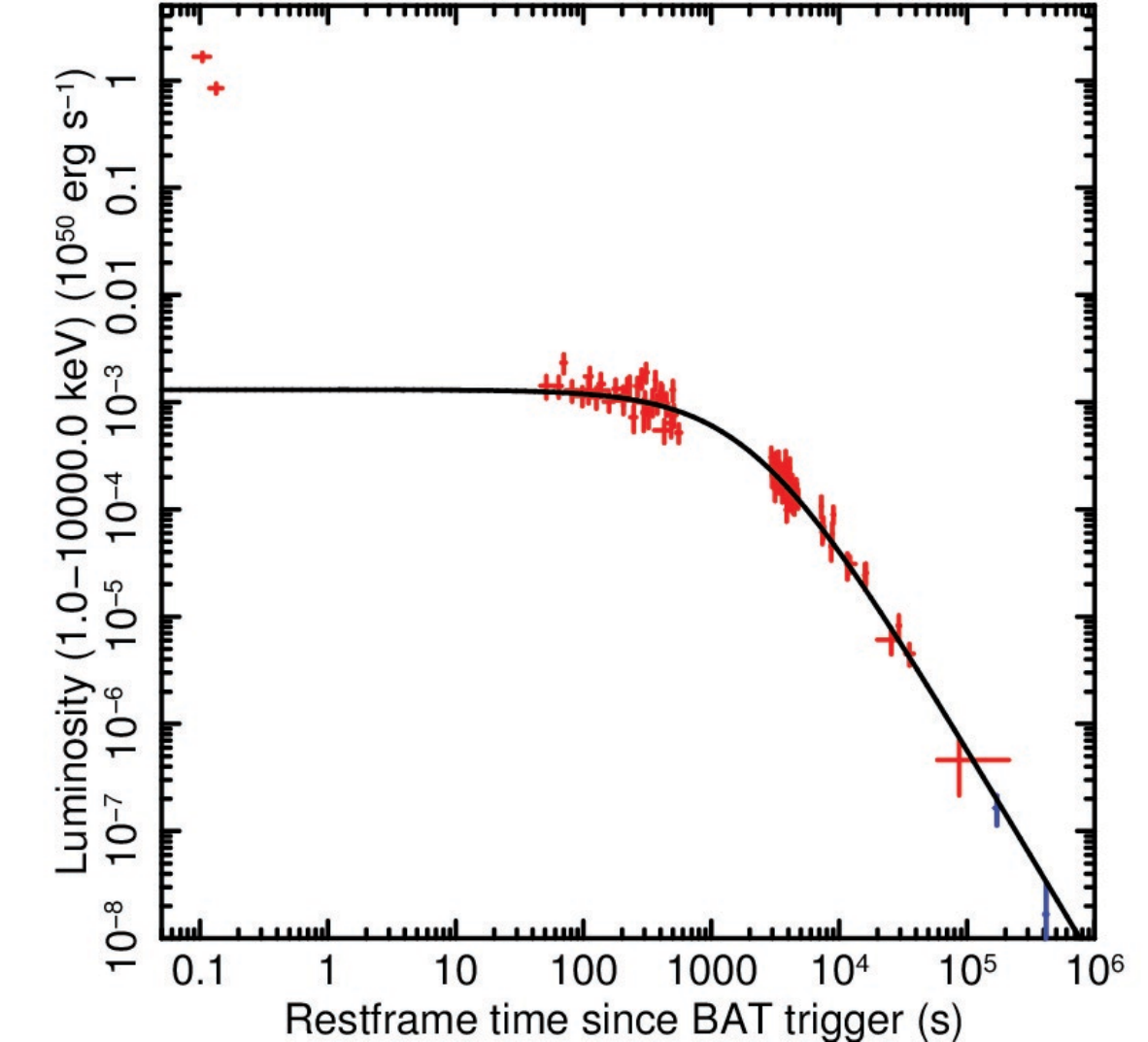}
\caption{Rest frame 1-10000 keV light curve for GRB 130603B. The red data are {\it Swift} observations, with the first 2 data points being from the BAT observations and the remainder are XRT observations. The blue data are the {\it XMM} observations obtained by \citet{fon13b}. The {\it Swift} XRT data are fitted using the magnetar model, as described in \citet{row13}, shown by the black line.
}
\label{magnetarfig}
\end{center}
\end{figure}

An alternative interpretation can be obtained by studying the origin of the early-time energy injection in more detail. In the standard progenitor theory for short GRBs, the compact binary merger of two neutron stars or a neutron star and a black hole will result in a black hole. The majority of accretion will occur within the first seconds \citep[e.g.][]{rez11}, which means that there is no large reservoir of material in an accretion disk typically used to explain the long-term energy injection. However, the merger of two neutron stars may instead form a magnetar with millisecond spin periods \citep{dai98} which can power a plateau phase as it spins down \citep{zha01}. This model has been fit to a large sample of short GRBs with evidence of energy injection and many were found to be consistent with the magnetar central engine model \citep{row13}. By fitting this model to GRB 130603B, we find that the X-ray light curve can be fitted with a newly-formed stable magnetar (with a magnetic field of $(8.59 \pm 0.19) \left( \frac{\epsilon}{1-\cos\theta} \right) ^{0.5} \times 10^{15}$ G and a spin period of $8.44^{+0.41}_{-0.39} \left( \frac{\epsilon}{1-\cos\theta} \right) ^{0.5}$ ms, where $\epsilon$ is the efficiency that the rotational energy is converted to X-rays and $\theta$ is the beaming angle of the magnetar emission, using the method described in \citealt{row13}). The magnetar model is shown in Figure \ref{magnetarfig} and assumes that the GRB is a ``naked'' GRB with no standard afterglow component. We also plot the two late time {\it XMM} observations obtained by \cite{fon13b}, these data are consistent with the magnetar fit obtained using the XRT data. As this model describes the entire light curve, the fireball model is no longer relevant and the possible jet break mentioned previously was a consequence of approximating a smooth curve with a broken power law model.

The early time optical data (before the 0.25 day break) are significantly fainter than expected from the X-ray observations and this discrepancy cannot be explained using absorption, as the absorption is well fitted using late time observations, or using the standard afterglow spectrum. The excess X-ray emission is consistent with that observed in many of the short GRBs previously fitted with the magnetar model, in which the discrepancy is explained as resulting from additional energy injection in the X-ray light curve \citep{row13}. Within the context of the magnetar model the late-time optical emission could be explained as reprocessing of the X-ray emission. For instance, short GRBs are expected to eject a small amount of material which produces a kilonova \citep[e.g.][]{li98, met10}, this kilonova component would be boosted by additional heating of the ejecta by the magnetar \citep{kul05}. Also the newly formed magnetar is expected to produce powerful winds, in some cases reaching relativistic velocities, which interact with the surrounding medium producing synchrotron emission peaking on the deceleration timescale of the magnetar \citep[approximately the plateau duration,][]{gao13}.

\end{appendix}

\end{document}